\documentclass[aps,prd,reprint,nofootinbib,superscriptaddress,floatfix]{revtex4-2}
\usepackage[T1]{fontenc}
\usepackage[utf8]{inputenc}
\usepackage{amsmath,amssymb,graphicx,bm}
\usepackage[hidelinks]{hyperref}
\hypersetup{pdftitle={Scalarized Einstein-Maxwell black holes in external scalar fields}}
\begin{document}
\title{Scalarized Einstein--Maxwell black holes in external scalar fields}

\author{Lang Cheng}
\email{langcheng@stu.scu.edu.cn}
\affiliation{College of Physics, Sichuan University, Chengdu 610065, China}
\author{Guangzhou Guo}
\email{guangzhou@sjtu.edu.cn}
\affiliation{Tsung-Dao Lee Institute, Shanghai Jiao Tong University, Shanghai 201210, China}
\author{Yuhan Li}
\email{leeyhan@stu.scu.edu.cn}
\affiliation{College of Physics, Sichuan University, Chengdu 610065, China}
\author{Peng Wang}
\email{pengw@scu.edu.cn}
\affiliation{College of Physics, Sichuan University, Chengdu 610065, China}

\begin{abstract}
We construct static black holes in
Einstein--Maxwell--scalar theory under prescribed external
scalar fields, mapping their existence domain and determining
how their scalar charges vary with the external field.
Focusing on solutions with positive scalar charge, we find
that a positive external field induces scalar charge in
Reissner–Nordström (RN) black holes below the zero-source scalarization threshold
and extends the allowed charge-to-mass range.
A negative external field instead introduces a finite lower
existence boundary and permits distinct black holes under identical external conditions.
Coupling these isolated responses through the leading companion-generated scalar fields, we find mutual scalar induction in binaries whose components are RN black holes in isolation. For binaries with scalar charges of the same sign, their scalar charges need not increase as the separation decreases, while some binaries with opposite scalar charges admit multiple equilibria and cannot continue along their original equilibrium branches to smaller separations.
\end{abstract}
\maketitle
\raggedbottom

\section{Introduction}
\label{sec:introduction}

Spontaneous scalarization allows black holes to acquire
scalar hair through an instability of a scalar-free solution,
even when the scalar field vanishes at
infinity~\cite{Damour:1993hw,
Damour:1996ke,
Harada:1998ge,
Gubser:2005ih,
Stefanov:2007eq,
Doneva:2010ke,
Cardoso:2013fwa,
Cardoso:2013opa,
Doneva:2022ewd}. Two well-studied realizations are Einstein--scalar--Gauss--Bonnet
(EsGB) gravity and Einstein--Maxwell--scalar (EMS) theory.
In EsGB gravity, the coupling to the Gauss--Bonnet invariant
supports hairy black holes and, for suitable coupling functions, triggers the
scalarization of Schwarzschild or Kerr black
holes~\cite{Doneva:2017bvd,Silva:2017uqg,Antoniou:2017acq,Antoniou:2017hxj,Doneva:2018rou,
Cunha:2019dwb,
Dima:2020yac,
Berti:2020kgk,
Doneva:2021dqn,
East:2021bqk}.
In EMS theory, a nonminimal coupling between the scalar
and electromagnetic fields allows scalarized branches to
bifurcate from Reissner--Nordstr\"om (RN) black
holes~\cite{Herdeiro:2018wub}.
In both theories, static scalar zero modes identify the
bifurcation thresholds, while nonlinear interactions
determine the properties of the resulting black hole
families~\cite{Blazquez-Salcedo:2018jnn,Silva:2018qhn,
Myung:2018jvi}.
Beyond the static problem, EMS theory is particularly
amenable to fully nonlinear evolutions because it
contains no higher-curvature terms.
Such evolutions have shown that unstable RN black holes
can develop scalar hair and settle into scalarized
configurations~\cite{Herdeiro:2018wub}.
Subsequent studies have investigated the existence domains,
stability, quasinormal-mode spectra, and potential
observational signatures of scalarized EMS black
holes~\cite{
Myung:2018jvi,
Fernandes:2019rez,
Myung:2019oua,
Konoplya:2019goy,
Fernandes:2019kmh,
Zou:2019bpt,
Blazquez-Salcedo:2020nhs,
Fernandes:2020gay,
LuisBlazquez-Salcedo:2020rqp,
Myung:2020dqt,
Myung:2020ctt,
Wang:2020ohb,
Guo:2021zed,
Gan:2021pwu,
Gan:2021xdl,
Guo:2021enm,
Guo:2022umh,
Guo:2022muy,
Lai:2022spn,
Hod:2022txa,
Chen:2022qrw,
Lai:2022ppn,
Guo:2022ghl,
Guo:2023ivz,
Guo:2023mda,
Belkhadria:2023ooc,
Chen:2023qic,
Xiong:2023bpl,
Chen:2024ilc,
Melis:2024kfr,
Zhang:2025jlb,
Cheng:2025hdw,
Wang:2025mxe,
Cheng:2026gxu,
Chen:2026rhr}.

Much of the early work on scalarized black holes focused
on isolated solutions with a vanishing asymptotic scalar field.
A black hole in a scalar environment, however, can experience
a nonzero background field that varies slowly across its
near-horizon region.
In a widely separated binary, a scalarized companion
provides such a background: its leading monopolar field
is approximately uniform near the other black hole.
This companion field can therefore be represented by
a nonzero asymptotic scalar value in the corresponding
isolated black hole problem, even if the scalar field
vanishes far from the binary~\cite{Khalil:2019wyy,Julie:2022huo}.
In a reversible, adiabatic description, the black hole's
response can then be studied by varying this background
at fixed entropy and, in EMS theory, fixed electric charge.
The associated mass variation is governed by the scalar-work
term in the first law~\cite{Gibbons:1996af,Khalil:2019wyy}.

The consequences of this environmental dependence have
been explored extensively in EsGB gravity.
Studies of black hole sensitivities have shown that
companion-generated fields can modify scalar hair and
drive black holes toward the boundaries of their
equilibrium branches~\cite{Julie:2022huo}.
Companion-induced dynamical scalarization has been studied
using adiabatic models and numerical
relativity~\cite{Julie:2023ncq,Capuano:2026lhs},
while numerical evolutions have also revealed
descalarization during mergers and scalar-charge reversals
in binaries with opposite scalar
charges~\cite{Silva:2020omi,Lara:2025kzj}.

In EMS theory, the influence of external scalar fields
has also been investigated for isolated and binary black
holes. Static solutions with nonzero asymptotic scalar
fields have been constructed to determine how the
black hole mass and scalar charge respond to the ambient
field at fixed entropy and electric charge.
Particular attention was paid to the vicinity of the RN
scalarization threshold, where the nonlinear response
was used to develop an effective description of
spontaneous and dynamical scalarization in
binaries~\cite{Khalil:2019wyy}.
Further studies have examined the existence domains
and scalar perturbations of EMS black holes with
nonzero asymptotic scalar
fields~\cite{Guo:2025xwh}.

Alongside these equilibrium studies, numerical evolutions
have explored the formation and loss of scalar hair in
EMS theory. They have demonstrated the growth and
saturation of scalar hair, as well as descalarization
through accretion or charge
extraction~\cite{Xiong:2022ozw,Niu:2022zlf,Corelli:2021ikv}.
Other coupling choices exhibit type I critical behavior
involving unstable intermediate black
holes~\cite{Zhang:2021nnn,Jiang:2023yyn}.
More recently, fully coupled head-on collisions of
initially unscalarized RN black holes have yielded
either scalarized or descalarized
remnants~\cite{Diaz:2026iyx}.
These early binary results motivate a systematic study
of how individual black holes respond to the scalar fields
generated by their companions, particularly beyond the
vicinity of the RN scalarization threshold.

Here we study static, spherically symmetric EMS black holes
with an exponential scalar--Maxwell coupling in external
scalar fields, extending the analysis beyond the immediate
vicinity of the RN scalarization threshold.
We map their existence domain and examine how their scalar
charges respond to the external field at fixed entropy
and electric charge, finding nonmonotonic behavior
and multiple equilibria under the same external field.
Using these isolated responses, we construct binary
equilibrium sequences that exhibit companion-induced
changes in scalar hair, multiple configurations, and
separation turning points for oppositely scalarized pairs.

The paper is organized as follows.
Sec.~\ref{sec:sourced} introduces the EMS model,
constructs black hole solutions in external scalar fields,
and examines their existence domain and equilibrium
sequences at fixed entropy and electric charge.
Sec.~\ref{sec:nonlinear-response} investigates the nonlinear
scalar response of isolated black holes and its implications
for binary equilibria, including the effects of the relative
signs of the two black holes' scalar charges and their
mass ratio.
Sec.~\ref{sec:discussion} discusses the main findings
and their possible dynamical implications.
The appendices provide details of the numerical construction,
the mass identity, and the near-threshold scaling behavior.

\section{Sourced scalarized black holes and their existence domain}
\label{sec:sourced}

In this section, we first describe static, electrically charged EMS black holes in an external scalar field and introduce their asymptotic charges and the scalar-work first law. We then examine how the external field changes their domain of existence and trace the equilibrium sequences at fixed entropy
and electric charge.

\subsection{Model and equilibrium black holes}
\label{sec:model}

Einstein--Maxwell--scalar theory allows electrically charged black holes to support scalar hair through a nonminimal coupling between the scalar and electromagnetic fields~\cite{Herdeiro:2018wub,Fernandes:2019rez}. We consider a real, massless scalar field minimally coupled to gravity, with its interaction with the electromagnetic field described by a coupling function \(f(\phi)\). In units \(c=G=1\), the action, with electromagnetic field strength \(F=d\mathcal A\), is
\begin{equation}
\begin{aligned}
I&=\frac{1}{16\pi}\int d^4x\,\sqrt{-g}\,
\bigl[R-2\nabla_\mu\phi\nabla^\mu\phi\\
&\hspace{28mm}-f(\phi)F_{\mu\nu}F^{\mu\nu}\bigr].
\end{aligned}
\label{eq:action}
\end{equation}
We adopt \(f(\phi)=e^{\alpha\phi^2}\), for which the scalar field equation becomes
\begin{equation}
\nabla_\mu\nabla^\mu\phi
=\frac14\frac{df}{d\phi}F_{\mu\nu}F^{\mu\nu}.
\label{eq:scalar-field}
\end{equation}
The even coupling makes the theory invariant under \(\phi\to-\phi\) and admits the RN solution with \(\phi=0\). For positive \(\alpha\), scalar zero modes can appear on the RN background, marking the onset of scalarized black hole solutions~\cite{Herdeiro:2018wub,Fernandes:2019rez,Myung:2018vug,Hod:2019ulh}.

In areal-radius coordinates, a static, spherically symmetric, electrically charged black hole is described by
\begin{equation}
\begin{aligned}
ds^2&=-N(r)e^{-2\delta(r)}dt^2+\frac{dr^2}{N(r)}+r^2d\Omega_2^2,\\
\mathcal A&=V(r)\,dt,\qquad \phi=\phi(r).
\end{aligned}
\label{eq:ansatz}
\end{equation}
A regular, nonextremal horizon is located at \(r=r_h\), where \(N(r_h)=0\) is a simple zero and \(\delta\), \(\phi\), and the electromagnetic field strength remain regular. The exterior region is static, with \(N(r)>0\) for \(r>r_h\). Asymptotic flatness requires \(N(\infty)=1\), and we normalize the time coordinate by setting \(\delta(\infty)=0\). The ADM mass \(M\) is determined by the asymptotic expansion \(N(r)=1-2M/r+O(r^{-2})\).

For the isolated black hole solutions, we choose
the conserved electric charge \(Q_e>0\) without loss
of generality, since reversing the Maxwell field
leaves the metric and scalar profile unchanged.
The charge is defined by the Maxwell flux,
\begin{equation}
Q_e=\frac{1}{4\pi}\int_{S^2_\infty} f(\phi)\,{}^\star F.
\label{eq:electric-charge}
\end{equation}
The Maxwell equation ensures that this flux is independent of the enclosing sphere. The Coulomb tail of the electric potential, however, is governed by \(Q_e/f(\phi_\infty)\), where \(\phi_\infty\) is the asymptotic scalar field. Thus, the Coulomb strength differs from the conserved electric charge when \(\phi_\infty\neq0\).

At large radii, the scalar field has the asymptotic expansion
\begin{equation}
\phi(r)=\phi_\infty+\frac{Q_s}{r}+O(r^{-2}).
\label{eq:scalar-asymptotics}
\end{equation}
The constant \(\phi_\infty\) specifies the external scalar field imposed on the black hole. Since the scalar has no potential, a nonzero asymptotic value is compatible with asymptotic flatness. The coefficient \(Q_s\) of the \(1/r\) term defines the scalar charge of each regular black hole solution. A nonzero scalar charge is not necessarily induced by the external field, since spontaneously scalarized black holes can carry \(Q_s\ne0\) even when \(\phi_\infty=0\). We use \(q=Q_e/M\) as the electric charge-to-mass ratio and \(Q_s/Q_e\) as the normalized scalar charge.

For variations among black hole solutions at fixed \(\alpha\), the first law takes the form~\cite{Gibbons:1996af,Khalil:2019wyy},
\begin{equation}
dM=T\,dS+\Phi_e\,dQ_e-Q_s\,d\phi_\infty.
\label{eq:scalar-work}
\end{equation}
Here \(S=A_H/4=\pi r_h^2\) is the black hole entropy, \(T\) is the Hawking temperature, and \(\Phi_e\) is the electric potential difference conjugate to the conserved charge \(Q_e\). The last term accounts for the work associated with changing the asymptotic scalar field. In particular, the ADM mass can vary with \(\phi_\infty\) even when \(S\) and \(Q_e\) are held fixed. The minus sign is consistent with our definition of \(Q_s\) in Eq.~\eqref{eq:scalar-asymptotics}.

We restrict attention to fundamental, nodeless solutions. When \(\phi_\infty\neq0\), the scalar field includes an imposed constant background, so radial nodes are counted in \(\phi(r)-\phi_\infty\), rather than in \(\phi(r)\) itself. Because the coupling \(f(\phi)\) is even, the transformation \(\phi(r)\to-\phi(r)\) reverses both \(\phi_\infty\) and \(Q_s\) while leaving the metric and electromagnetic field unchanged. We therefore follow solutions that approach either RN or a scalarized black hole with \(Q_s>0\) as \(\phi_\infty\to0\), without displaying their reflected counterparts separately.

For prescribed \((S,Q_e,\alpha,\phi_\infty)\), we construct static black hole solutions by solving the radial field equations using Chebyshev pseudospectral collocation on a compactified domain and Newton iteration. The ADM mass \(M\) and scalar charge \(Q_s\) are obtained from the asymptotic fields. Varying \(\phi_\infty\) and \(Q_e\) allows us to map the existence domain of the black holes studied here. Independent shooting calculations provide additional checks. Details of the numerical construction, including continuation through turning points in $\phi_\infty$, are given in Appendix~\ref{app:spectral}, while Appendix~\ref{app:identities} presents the extraction of physical observables and the derivation of the mass identity.

\subsection{Existence domain}
\label{sec:existence}

\begin{figure*}[tp]
\centering
\includegraphics[width=\textwidth]{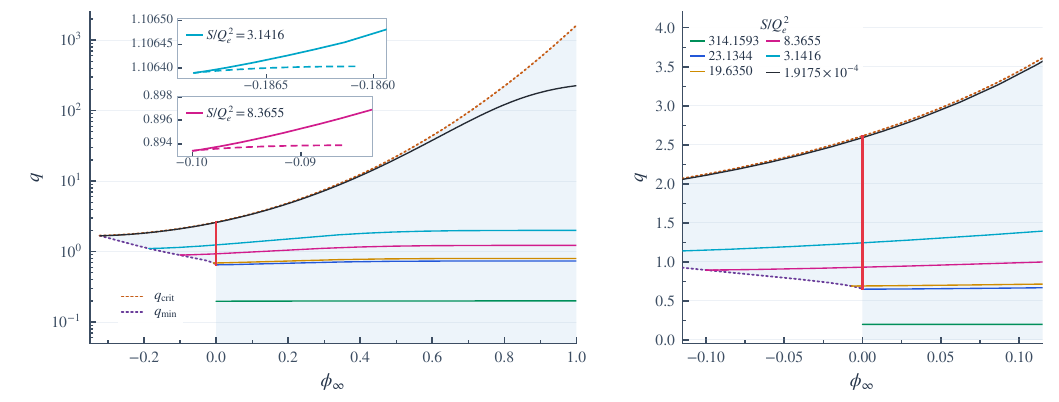}
\caption{Existence domain of scalarized EMS black holes in an external scalar field at \(\alpha=10\), shown in the \((\phi_\infty,q)\) plane with \(q=Q_e/M\). The left panel shows the broad domain, while the right panel enlarges the region near zero source. At \(\phi_\infty=0\), the red vertical segment represents scalarized black holes extending from the RN bifurcation to the singular critical solution. For nonzero source, the upper boundary is the critical line \(q_{\rm crit}(\phi_\infty)\) (orange dotted), which continues the zero-source critical endpoint. The lower boundary is \(q_{\min}=0\) for positive source, corresponding to the constant-scalar Schwarzschild limit, and \(q_{\min}(\phi_\infty)>0\) (purple dotted) for negative source. The two boundaries meet at \(\phi_{\infty,\min}\simeq-0.3234\). Colored curves represent equilibrium sequences at fixed \(S\) and \(Q_e\), labeled by \(S/Q_e^2\). At negative source, solid and dashed curves denote the decreasing and returning branches, respectively. The insets magnify the small separation in \(q\) between two distinct black holes at the same \(\phi_\infty\) for the \(S/Q_e^2=3.1416\) and \(8.3655\) sequences.}

\label{fig:1}
\end{figure*}

We examine how an external scalar field changes the existence domain of static black holes, taking \(\alpha=10\) throughout the numerical analysis. For the displayed solutions with \(Q_s>0\), positive and negative sources correspond to \(\phi_\infty>0\) and \(\phi_\infty<0\), respectively. Fig.~\ref{fig:1} shows the existence domain, with an overview in the left panel and an enlargement near \(\phi_\infty=0\) in the right panel. At zero source, the red vertical segment represents scalarized black holes extending from their bifurcation with RN at \(q=q_{\mathrm{bif}}\) to the singular critical solution known from the zero-source EMS model~\cite{Herdeiro:2018wub,Fernandes:2019rez}.

To understand the lower boundary of the existence domain for \(\phi_\infty>0\), we consider the scalar deformation induced by a weak electric field. At fixed \(S\), \(\alpha\), and \(\phi_\infty\), we expand around the uncharged solution as
\begin{equation}
\begin{aligned}
\phi(r)&=\phi_\infty+Q_e^2\psi(r)+O(Q_e^4),\\
N(r)&=1-\frac{r_h}{r}+O(Q_e^2),\qquad \delta(r)=O(Q_e^4).
\end{aligned}
\label{eq:weak-expansion}
\end{equation}
Since the electromagnetic source in Eq.~\eqref{eq:scalar-field} is quadratic in \(Q_e\), the scalar deformation begins at order \(Q_e^2\). At this order, the scalar field equation reduces to
\begin{equation}
\left[r(r-r_h)\psi'(r)\right]'
=-\frac{\alpha\phi_\infty e^{-\alpha\phi_\infty^2}}{r^2},
\label{eq:weak-scalar-ode}
\end{equation}
where a prime denotes \(d/dr\). Regularity at the horizon eliminates the integration constant that would otherwise produce a logarithmic divergence, while \(\psi(\infty)=0\) fixes the remaining additive constant. The resulting scalar profile is
\begin{equation}
\psi(r)=\frac{\alpha\phi_\infty e^{-\alpha\phi_\infty^2}}{r_h r}.
\label{eq:weak-scalar-solution}
\end{equation}

Comparing the \(1/r\) coefficient with Eq.~\eqref{eq:scalar-asymptotics} and using \(S=\pi r_h^2\), we obtain
\begin{equation}
\frac{Q_s}{Q_e}
=\alpha\phi_\infty e^{-\alpha\phi_\infty^2}\sqrt{\frac{\pi}{S}}\,Q_e
+O\!\left[\left(\sqrt{\frac{\pi}{S}}\,Q_e\right)^3\right].
\label{eq:weak-response}
\end{equation}
This leading-order result holds for \(\sqrt{\pi/S}\,Q_e\ll1\). For \(\alpha>0\) and \(\phi_\infty>0\), even an arbitrarily small positive electric charge induces \(Q_s>0\). As \(Q_e\to0\), the mass approaches \(M=r_h/2\), so \(q=Q_e/M\to0\) and the lower boundary is \(q_{\min}=0\). At exactly \(Q_e=0\), the limiting solution is Schwarzschild with constant \(\phi=\phi_\infty\).

For each imposed \(\phi_\infty\), the upper boundary of the existence domain is defined by a critical charge-to-mass ratio \(q=q_{\rm crit}(\phi_\infty)\). As shown in Fig.~\ref{fig:1}, a positive external scalar field shifts this boundary to larger \(q\), allowing black holes to attain charge-to-mass ratios beyond the zero-source limit. As the critical line is approached at fixed \(Q_e\), the horizon area tends to zero while the ADM mass remains finite and the Hawking temperature increases. This critical line extends the singular critical endpoint of the zero-source scalarized solutions to nonzero external fields~\cite{Herdeiro:2018wub}. The limiting configuration is therefore singular rather than a regular extremal black hole.

For negative external fields, the black holes considered here remain bounded above by \(q_{\rm crit}(\phi_\infty)\), but a finite lower boundary \(q_{\min}(\phi_\infty)>0\) also appears. The weak-charge expansion in Eq.~\eqref{eq:weak-response} gives \(Q_s<0\) for \(\phi_\infty<0\), so the \(Q_s>0\) solutions cannot extend to the weak-charge Schwarzschild limit. Instead, they retain nonzero scalar charge at their lower boundary. As shown in Fig.~\ref{fig:1}, making the external field more negative raises \(q_{\min}\) and lowers \(q_{\rm crit}\), progressively narrowing the allowed range of \(q\). The two boundaries meet at the numerically estimated endpoint \(\phi_{\infty,\min}\simeq-0.3234\), where \(q\simeq1.6869\). This marks the most negative external scalar field for which the fundamental \(Q_s>0\) black holes studied here exist.

Near the RN bifurcation, a weak negative external scalar field raises the lower boundary above \(q_{\mathrm{bif}}\). The local effective description of scalarization~\cite{Khalil:2019wyy} predicts the following scaling as \(\phi_\infty\to0^-\):
\begin{equation}
q_{\min}(\phi_\infty)-q_{\mathrm{bif}}=C_{\mathrm{ZM}}(-\phi_\infty)^{2/3}+O\!\left((-\phi_\infty)^{4/3}\right).
\label{eq:boundary-scaling}
\end{equation}
Appendix~\ref{app:zero-mode} derives this two-thirds-power scaling, determines \(C_{\mathrm{ZM}}\), and compares it with the numerical lower boundary.

The colored curves in Fig.~\ref{fig:1} show how the equilibrium black hole solutions change as \(\phi_\infty\) is varied at fixed \(S\) and \(Q_e\). Each curve is continued from a black hole at \(\phi_\infty=0\), and the labels \(S/Q_e^2\) distinguish the different sequences. Holding \(S\) fixed corresponds to the reversible, adiabatic equilibrium approximation~\cite{Cardenas:2017chu,Khalil:2019wyy}. The electric charge is also held fixed, since the scalar is electrically neutral and carries no Maxwell charge in the absence of charged matter flux. The first law, Eq.~\eqref{eq:scalar-work}, then gives \((\partial M/\partial\phi_\infty)_{S,Q_e,\alpha}=-Q_s\). Thus, the ADM mass, and consequently \(q=Q_e/M\), can vary along each sequence even though \(S\) and \(Q_e\) remain fixed.

The fixed-\((S,Q_e)\) sequences in Fig.~\ref{fig:1} begin at zero source, either from RN black holes below the scalarization bifurcation or from scalarized black holes above it. Along the \(S/Q_e^2=314.1593\) sequence, a positive external field induces scalar hair on an initially RN black hole, which disappears as \(\phi_\infty\to0\). The \(S/Q_e^2=23.1344\) sequence starts at the marginal RN solution at \(q=q_{\mathrm{bif}}\), separating these two cases. By contrast, sequences starting from scalarized black holes already carry nonzero scalar hair at zero source and extend smoothly toward the large positive external fields shown in Fig.~\ref{fig:1}.

The negative-source continuation of these initially scalarized black holes is more intricate. As \(\phi_\infty\) is decreased from zero at fixed \((S,Q_e)\), the sequence reaches its most negative external field, where it meets the lower existence boundary \(q_{\min}(\phi_\infty)\). We call the segment leading to this point the decreasing branch. Continuing past this turning point, \(\phi_\infty\) increases again along the returning branch. Thus, for values of \(\phi_\infty\) slightly above the minimum, the two branches describe distinct black holes with the same \(S\), \(Q_e\), and \(\phi_\infty\). Their scalar charges and ADM masses differ, although the corresponding values of \(q\) are very close. At the minimum itself, the two branches meet in a single solution. The two insets of Fig.~\ref{fig:1} illustrate this behavior for \(S/Q_e^2=3.1416\) and \(8.3655\), revealing the small separation between the branches that is nearly invisible in the broader view.

\section{Scalar Response and Binary Equilibria}
\label{sec:nonlinear-response}

An external scalar field can modify the scalar charge of a black hole, inducing scalar hair or changing its existing scalar configuration. We first examine this response for isolated black holes at fixed entropy and electric charge, focusing on how the scalar charge varies with the external field. We then turn to binary systems, where each black hole generates the external field acting on its companion. Their mutual response gives rise to equilibrium configurations whose properties depend on the scalar charges and mass ratio of the two black holes.

\subsection{Response to a prescribed scalar field}
\label{sec:response}

\begin{figure*}[tp]
\centering
\includegraphics[width=\textwidth]{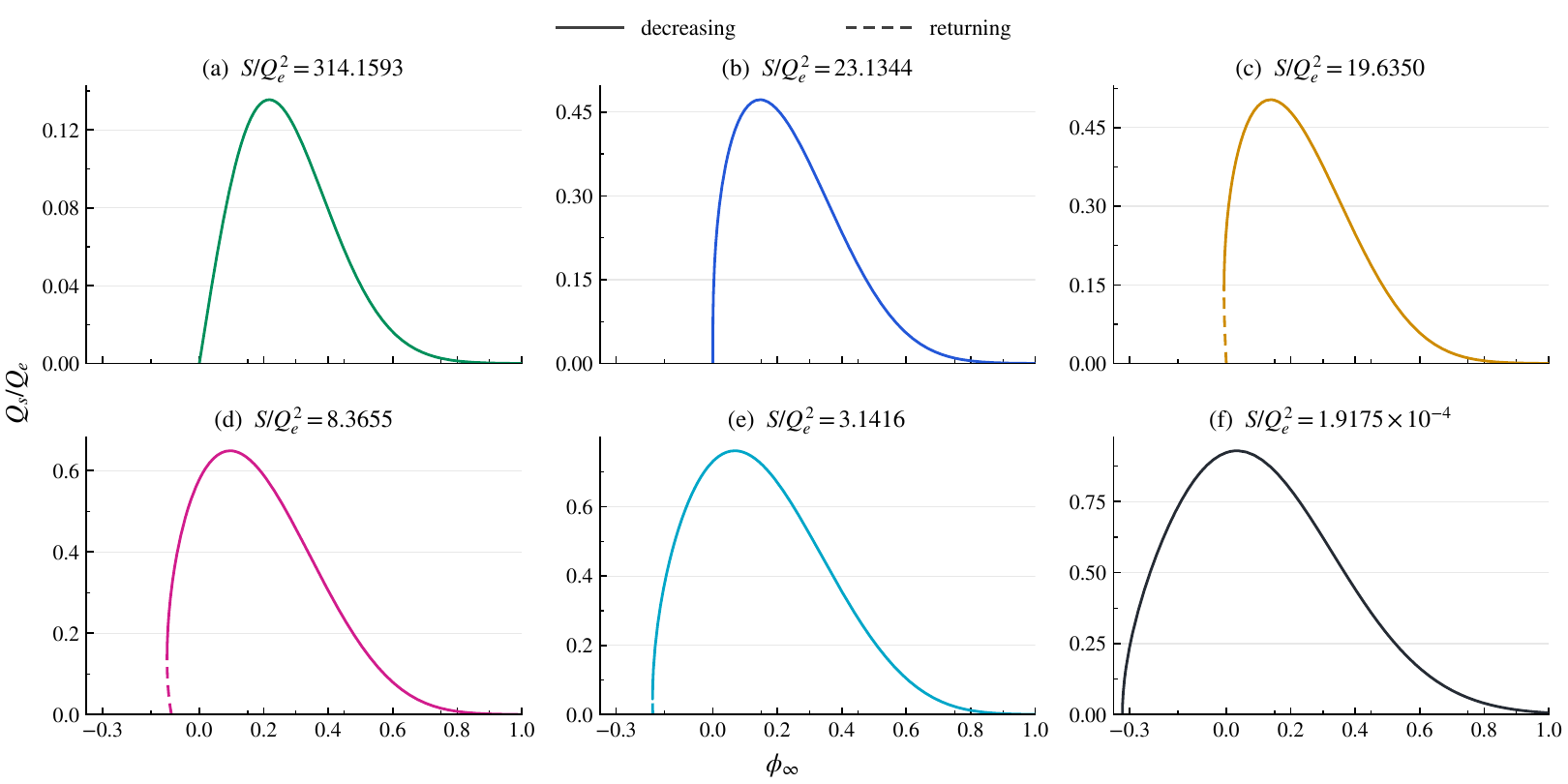}
\caption{Charge-normalized scalar response \(Q_s/Q_e\) versus the imposed field \(\phi_\infty\) along fixed-\((S,Q_e)\) equilibrium sequences, with \(\alpha=10\). The panel labels give \(S/Q_e^2\), and the colors match the corresponding families in Fig.~\ref{fig:1}. Panel (a) shows a weakly charged RN black hole, panel (b) the marginal RN solution at the scalarization threshold, and panels (c)--(f) black holes that are already scalarized at zero source. For positive source, the scalar charge generally grows, reaches a maximum, and then decreases. For families that start from zero-source scalarized black holes, negative source suppresses the scalar charge and produces decreasing and returning branches, shown as solid and dashed curves, respectively.}
\label{fig:2}
\end{figure*}

\begin{figure*}[tp]
\centering
\includegraphics[width=\textwidth]{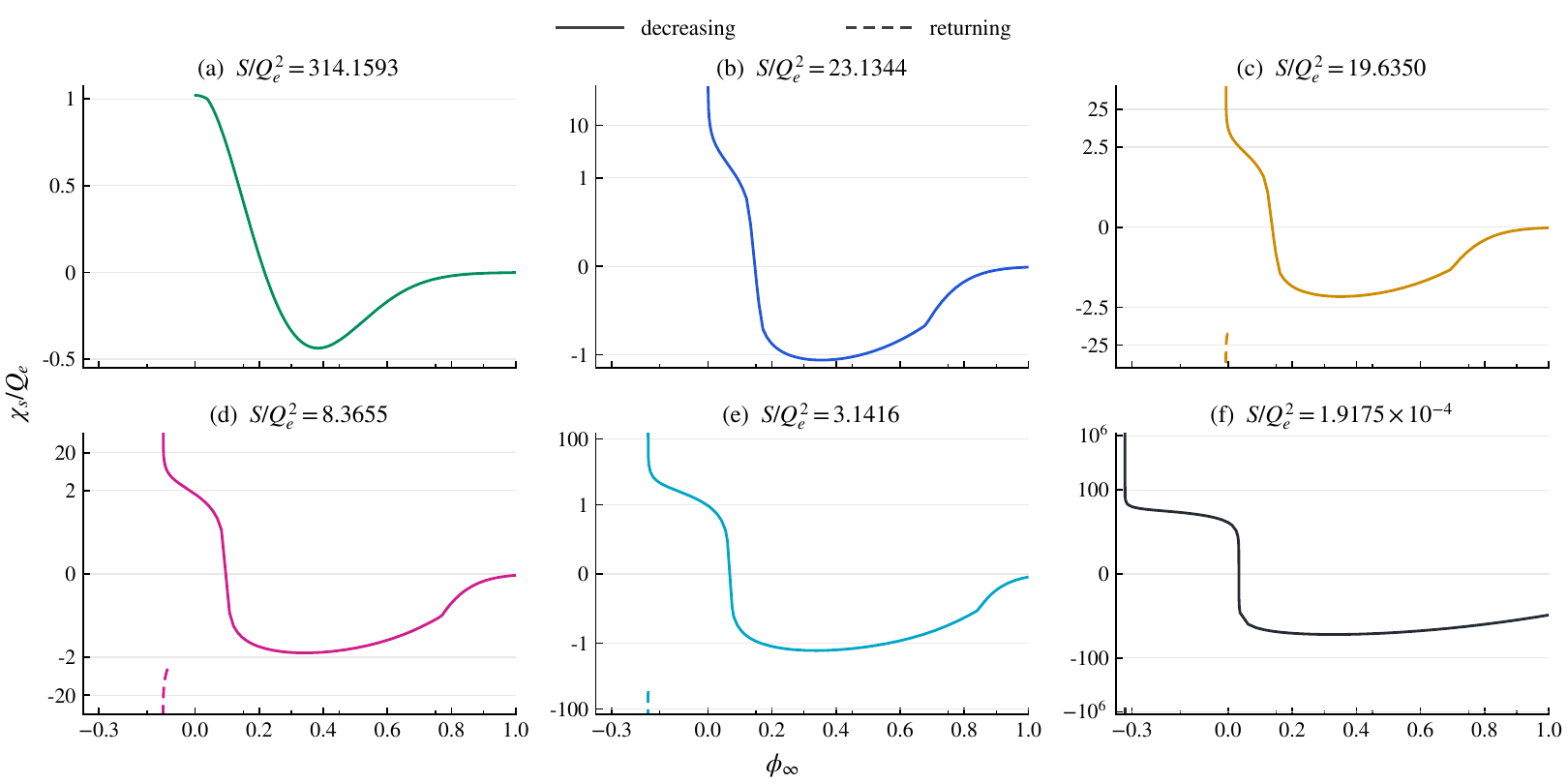}

\caption{Charge-normalized scalar susceptibility \(\chi_s/Q_e\) for the equilibrium sequences in Fig.~\ref{fig:2}, with the same panel labels, colors, and line styles. The susceptibility is defined in Eq.~\eqref{eq:susceptibility}. All panels use a symmetric logarithmic vertical scale, which is linear near zero and logarithmic for larger positive and negative values. Positive and negative values indicate whether the scalar charge increases or decreases with the external field, while \(\chi_s=0\) marks a stationary point of \(Q_s\). The susceptibility diverges at the RN scalarization threshold as \(\phi_\infty\to0^+\). For sequences with a minimum external field, \(\chi_s\) diverges to \(+\infty\) on the decreasing branch and to \(-\infty\) on the returning branch as both approach the minimum source from above.}
\label{fig:3}
\end{figure*}

Along the fixed-\((S,Q_e)\) equilibrium sequences introduced in Sec.~\ref{sec:sourced}, the scalar charge \(Q_s\) varies with the imposed field \(\phi_\infty\). Following Ref.~\cite{Khalil:2019wyy}, we characterize this response by the scalar susceptibility
\begin{equation}
\chi_s=\left(\frac{\partial Q_s}{\partial\phi_\infty}\right)_{S,Q_e,\alpha}.
\label{eq:susceptibility}
\end{equation}
The sign of \(\chi_s\) indicates whether \(Q_s\) increases or decreases with \(\phi_\infty\), while \(\chi_s=0\) marks a stationary point of the scalar charge. Figs.~\ref{fig:2} and~\ref{fig:3} show \(Q_s/Q_e\) and \(\chi_s/Q_e\), respectively, for the same equilibrium sequences, ordered by decreasing \(S/Q_e^2\). Their zero-source starting points range from weakly charged RN black holes, through the RN scalarization threshold, to black holes already carrying scalar hair.

The sequence in Figs.~\ref{fig:2}(a) and~\ref{fig:3}(a) begins at a weakly charged RN black hole with \(Q_s=0\) at zero source. Since \(S/Q_e^2\) is large, the leading weak-charge result in Eq.~\eqref{eq:weak-response} describes the scalar response,
\begin{equation}
\frac{Q_s}{Q_e}\simeq
\alpha\sqrt{\pi}(S/Q_e^2)^{-1/2}
\phi_\infty e^{-\alpha\phi_\infty^2}.
\label{eq:weak-rn-response}
\end{equation}
As \(\phi_\infty\) increases from zero, the linear factor initially drives the growth of the scalar charge, while the exponential eventually suppresses it. In this approximation, the response reaches a finite maximum before tending back to zero as \(\phi_\infty\to+\infty\). Differentiating gives
\begin{equation}
\frac{\chi_s}{Q_e}\simeq
\alpha\sqrt{\pi}(S/Q_e^2)^{-1/2}
e^{-\alpha\phi_\infty^2}(1-2\alpha\phi_\infty^2).
\label{eq:weak-rn-susceptibility}
\end{equation}
The maximum occurs at \(\phi_\infty^{\rm max}=1/\sqrt{2\alpha}\simeq0.224\) for \(\alpha=10\), with height \((Q_s/Q_e)_{\rm max}\simeq\sqrt{\alpha\pi/(2e)}(S/Q_e^2)^{-1/2}\), which scales linearly with \(Q_e/\sqrt S\). Accordingly, the susceptibility is positive before the maximum, vanishes at it, and becomes negative beyond it, as shown in Fig.~\ref{fig:3}(a).

At the RN scalarization threshold, Figs.~\ref{fig:2}(b) and~\ref{fig:3}(b) show a continuous scalar response with a divergent slope at zero source. This marginal RN solution separates the RN and scalarized black holes at \(\phi_\infty=0\). At the bifurcation, the local cubic source relation of Ref.~\cite{Khalil:2019wyy} and Appendix~\ref{app:zero-mode} reduces to \(\phi_\infty\propto Q_s^3\), giving
\begin{equation}
Q_s\propto\phi_\infty^{1/3},
\qquad
\chi_s\propto\phi_\infty^{-2/3},
\qquad (\phi_\infty\to0^+).
\label{eq:rn-onset-scaling}
\end{equation}
Thus, an arbitrarily weak positive external field produces a much stronger scalar response than in the weakly charged RN regime, even though \(Q_s\) still tends continuously to zero. At larger positive source, the scalar charge again reaches a maximum and then decreases.

The sequences in Figs.~\ref{fig:2}(c)--(f) and~\ref{fig:3}(c)--(f) start from black holes that already carry scalar hair at \(\phi_\infty=0\). As the external field increases, their scalar charge initially grows, reaches a maximum, and then decreases toward small values. This nonmonotonic behavior resembles that of weakly charged RN black holes, but here the scalar charge remains nonzero even in the absence of an external field. Correspondingly, the susceptibility changes from positive to negative as the scalar charge passes through its maximum. As \(S/Q_e^2\) decreases, the maximum of \(Q_s/Q_e\) increases and shifts to smaller positive \(\phi_\infty\). However, its growth is much slower than the weak-charge scaling would suggest.

The weak-charge approximation predicts a peak height proportional to \(Q_e/\sqrt S\), but the full nonlinear solutions are subject to an exact constraint. The static spherical EMS equations satisfy the general mass relation~\cite{Gibbons:1996af,Guo:2026fga}
\begin{equation}
\left(\frac{M}{Q_e}\right)^2+
\left(\frac{Q_s}{Q_e}\right)^2
=e^{-\alpha\phi_\infty^2}+\left(\frac{2TS}{Q_e}\right)^2,
\label{eq:amplitude-balance}
\end{equation}
where \(T\) is the Hawking temperature. This identity follows by evaluating the EMS radial first integral derived in Ref.~\cite{Guo:2026fga} at the horizon and infinity. We reproduce the derivation in our conventions in Appendix~\ref{app:identities}. Its zero-source form was obtained in Ref.~\cite{Astefanesei:2019pfq}. Together with the Smarr relation \(M=2TS+\Phi_eQ_e\) and \(\Phi_eQ_e>0\), it gives
\begin{equation}
\left|\frac{Q_s}{Q_e}\right|
<e^{-\alpha\phi_\infty^2/2}\leq1.
\label{eq:scalar-charge-bound}
\end{equation}
The normalized scalar charge is therefore bounded, explaining why the weak-charge growth cannot persist as the electric charge becomes larger relative to the black hole entropy.

For sequences starting from scalarized black holes at zero source, a negative external field suppresses the scalar charge. This differs from the counter-scalarization associated with changes in the coupling~\cite{Belkhadria:2023ooc}, since \(\alpha\) is held fixed here. As \(\phi_\infty\) decreases along the solid decreasing branch, the sequence reaches its most negative external field. Continuing past this turning point leads to the dashed returning branch, along which \(\phi_\infty\) increases again. At the turning point, \((\partial\phi_\infty/\partial Q_s)_{S,Q_e,\alpha}=0\), and the susceptibility diverges, with \(\chi_s\to+\infty\) on the decreasing branch and \(\chi_s\to-\infty\) on the returning branch, as shown in Fig.~\ref{fig:3}. Some returning branches are too narrow to resolve on the common source scale, particularly in Fig.~\ref{fig:2}(f).

The first law also explains why the two branches are more easily distinguished in Figs.~\ref{fig:2} and~\ref{fig:3} than in the existence diagram of Fig.~\ref{fig:1}. Along a fixed-\((S,Q_e)\) sequence, \(dM=-Q_s\,d\phi_\infty\), which gives
\begin{equation}
\left(\frac{\partial q}{\partial\phi_\infty}\right)_{S,Q_e,\alpha}
=\frac{Q_eQ_s}{M^2}.
\label{eq:q-source-slope}
\end{equation}
The same thermodynamic relation also yields
\begin{equation}
\chi_s=-\left(\frac{\partial^2M}{\partial\phi_\infty^2}\right)_{S,Q_e,\alpha}.
\label{eq:susceptibility-mass}
\end{equation}
We have verified that the susceptibility obtained from the second derivative of the ADM mass agrees with that calculated directly from the scalar charge. At the minimum-source turning point, both \(M\) and \(Q_s\) remain finite, so the slope of \(q\) remains finite even though \(\chi_s\) diverges. Consequently, near the turning point, the decreasing and returning branches can have noticeably different scalar charges at the same external field, while their masses and values of \(q\) remain very close.

\subsection{Response to the scalar field of a companion}
\label{sec:binary-response}

In a sufficiently separated binary, the scalar field produced by each black hole acts as an external field on its companion. The companion responds by changing its scalar charge, which in turn modifies the field experienced by the first black hole. We describe this mutual response using the isolated equilibrium solutions of Sec.~\ref{sec:response}, keeping \((S,Q_{e})\) fixed for each black hole.

The companion's scalar field changes not only the scalar charge but also the mass of each black hole. At fixed \(S\) and \(Q_e\), the first law, Eq.~\eqref{eq:scalar-work}, reduces to \(dM=-Q_s\,d\phi_\infty\). This mass response is characterized by the black hole sensitivities~\cite{Khalil:2019wyy},
\begin{equation}
\begin{aligned}
\alpha_{\rm BH}&\equiv\left(\frac{\partial\ln M}{\partial\phi_\infty}\right)_{S,Q_e}=-\frac{Q_s}{M},\\
\beta_{\rm BH}&\equiv\left(\frac{\partial\alpha_{\rm BH}}{\partial\phi_\infty}\right)_{S,Q_e}
=-\frac{\chi_s}{M}-\left(\frac{Q_s}{M}\right)^2.
\end{aligned}
\label{eq:binary-sensitivities}
\end{equation}
Here \(M\) is the ADM mass and \(\chi_s\) is the scalar susceptibility defined in Eq.~\eqref{eq:susceptibility}. The isolated EMS response functions \(M(\phi_\infty)\), \(Q_s(\phi_\infty)\), and \(\chi_s(\phi_\infty)\) already incorporate the effects of each black hole's own electromagnetic field.
Black hole sensitivities have also been studied in EsGB gravity~\cite{Julie:2019sab,Julie:2022huo}, together with their implications for scalarization during binary inspiral~\cite{Julie:2023ncq}. Although the isolated solutions are constructed with positive electric charge, reversing its sign leaves the black hole mass and scalar charge, as well as their dependence on the external scalar field, unchanged. The two black holes in a binary may therefore carry electric charges of the same or opposite signs, independently of the signs of their scalar charges.

At large separation, each black hole experiences the
scalar field of its companion as an approximately
uniform background.
Assuming that the scalar field vanishes at spatial
infinity of the binary, the asymptotic behavior
in Eq.~\eqref{eq:scalar-asymptotics}
gives~\cite{Khalil:2019wyy}
\begin{equation}
\phi^A=\frac{Q_s^B(\phi^B)}{d},\qquad
\phi^B=\frac{Q_s^A(\phi^A)}{d}.
\label{eq:binary-feedback}
\end{equation}
Here \(d\) is the separation, while \(\phi^A\)
and \(\phi^B\) are the scalar fields produced by
B at A and by A at B, respectively.
The functions \(Q_s^A(\phi^A)\) and
\(Q_s^B(\phi^B)\) give the scalar charges of
A and B determined from their isolated responses
at fixed entropy and electric charge.
Eq.~\eqref{eq:binary-feedback} retains the
leading scalar monopole field of each companion,
proportional to \(1/d\).
Across the region near each black hole,
the companion's scalar field varies by a fraction
of order \(r_{h}^{i}/d\) relative to its
approximately uniform contribution \(Q_{s}^{j}/d\).
Subleading terms in the scalar field, the
companion's electromagnetic and gravitational
interactions, and orbital effects are not
included in this approximation.

The coupled scalar response equations in
Eq.~\eqref{eq:binary-feedback} are independent of
the relative signs of the electric charges,
whereas the orbital interaction is not.
At large separation, the leading interaction
potential is~\cite{Khalil:2019wyy}
\begin{equation}
U_{\mathrm{int}}(d)=
-\frac{
M_0^A M_0^B
+Q_s^A(0)Q_s^B(0)
-Q_{e}^{A}Q_{e}^{B}
}{d}.
\label{eq:binary-interaction-potential}
\end{equation}
Since \(|Q_{s}^{i}(0)|<|Q_{e}^{i}|\) for each isolated
black hole, electric charges of opposite signs
always give a net attractive interaction at this
order, regardless of the signs of the scalar charges.
When the electric charges have the same sign,
the interaction can be attractive or repulsive.
In particular, it is repulsive for the F3+F4
and F4+F4 configurations with opposite scalar
charges considered below.

\begin{table}[ht]
\caption{Four representative black hole response families at \(\alpha=10\) selected for the binary analysis. The table lists the properties of their zero-source (\(\phi_\infty=0\)) black holes, from which the equilibrium sequences are continued. The corresponding scalar responses are shown in the indicated panels of Fig.~\ref{fig:2}. Here \(q_0=|Q_{e}|/M_{0}\) is the magnitude of
the electric charge-to-mass ratio at zero source.}
\label{tab:binary-families}
\centering
\small
\setlength{\tabcolsep}{4pt}
\begin{tabular}{@{}lclcc@{}}
\hline\hline
Family & Fig.~\ref{fig:2} & Zero-source state & $q_0$ & $S/Q_e^2$ \\
\hline
F1 & (a) & RN & $0.1980$ & $314.1593$ \\
F2 & (b) & RN at onset & $0.6489$ & $23.1344$ \\
F3 & (d) & Scalarized & $0.9302$ & $8.3655$ \\
F4 & (e) & Scalarized & $1.2427$ & $3.1416$ \\
\hline\hline
\end{tabular}
\end{table}

To compare binaries whose components originate from RN or scalarized black holes, we select four representative black hole response families, F1--F4, whose zero-source members are listed in Tab.~\ref{tab:binary-families}. F1 starts from an RN black hole below the scalarization threshold, F2 from the marginal RN solution at the threshold, and F3 and F4 from scalarized black holes carrying nonzero scalar charge at \(\phi_\infty=0\). Their responses to an external scalar field are shown in the corresponding panels of Fig.~\ref{fig:2}. For each binary component \(i=A,B\), we denote the mass of its zero-source reference black hole by \(M_{0}^{i}\) and define \(M_0=M_0^A+M_0^B\). We normalize the binary separation and scalar charges as \(d/M_0\) and \(Q_{s}^{i}/M_{0}^{i}\), respectively. The zero-source mass ratio is
\begin{equation}
\mu=\frac{M_0^B}{M_0^A}.
\label{eq:binary-normalization}
\end{equation}

\begin{figure*}[t]
\centering
\includegraphics[width=\textwidth]{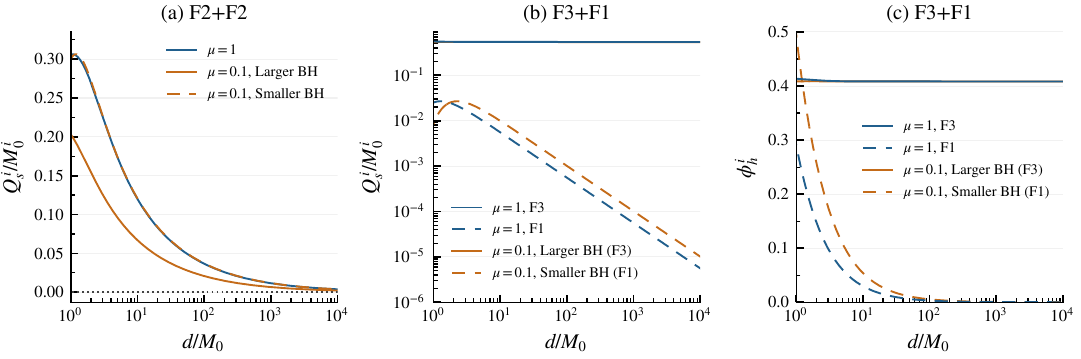}
\caption{Scalar responses of F2+F2 and F3+F1 binaries as functions of the separation \(d/M_0\). Blue and orange curves correspond to mass ratios \(\mu=1\) and \(0.1\), respectively. (a) F2+F2 binaries, whose components are RN black holes at the scalarization threshold when isolated. The dotted line represents the solution with vanishing scalar charges, while the colored curves show nonzero charges sustained by mutual induction. For \(\mu=1\), both components follow the same curve. For \(\mu=0.1\), solid and dashed orange curves represent the larger and smaller black holes, respectively. (b,c) F3+F1 binaries, with F3 already scalarized and F1 RN at zero source. Panel (b) shows their normalized scalar charges \(Q_{s}^{i}/M_{0}^{i}\), while panel (c) shows the horizon scalar fields \(\phi_h^i\). Solid and dashed curves denote F3 and F1, respectively. As the separation decreases, the induced scalar charge of F1 reaches a maximum and then declines, even as its horizon scalar field continues to increase. Here \(M_{0}^{i}\) is the zero-source mass of each component, and \(M_0=M_0^A+M_0^B\).}
\label{fig:binary-B1}
\end{figure*}

When both black holes are RN in isolation, \(Q_s^A(0)=Q_s^B(0)=0\), so Eq.~\eqref{eq:binary-feedback} always admits the solution \(\phi^A=\phi^B=0\), in which neither black hole carries scalar hair. However, the same equations can also admit equilibrium solutions with nonzero scalar charges. In such a solution, the scalar field produced by each black hole induces a scalar charge on its companion, which in turn sustains the field acting back on the first. This mutual induction can therefore support scalar hair even though neither black hole carries it in isolation.

For RN black holes with finite zero-source susceptibilities, the onset of mutual induction can be determined by linearizing Eq.~\eqref{eq:binary-feedback} around the zero-scalar solution. Using \(Q_s^A\simeq\chi_s^A(0)\phi^A\) and \(Q_s^B\simeq\chi_s^B(0)\phi^B\), we obtain \(d^2=\chi_s^A(0)\chi_s^B(0)\) as the threshold for a nonzero infinitesimal scalar configuration. At smaller separations, \(d^2<\chi_s^A(0)\chi_s^B(0)\), mutual induction can support finite scalar charges. The relation \(\phi^A/\phi^B=\chi_s^B(0)/d\) also shows that the positive susceptibilities of the RN black holes considered here favor induced scalar charges of the same sign, reinforcing each other's scalar fields.

The F2+F2 binary in Fig.~\ref{fig:binary-B1}(a) provides a different case. Both components are RN black holes at the scalarization threshold (Tab.~\ref{tab:binary-families}), where the zero-source susceptibility diverges. The finite-susceptibility criterion therefore no longer applies. Instead, the critical response \(Q_s\propto\phi_\infty^{1/3}\) discussed in Sec.~\ref{sec:response} allows mutual induction to sustain small nonzero scalar charges even at large separation. For equal masses (\(\mu=1\)), combining this response with Eq.~\eqref{eq:binary-feedback} gives \(Q_s\propto d^{-1/2}\) at large \(d\). The blue curve represents both identical black holes, whose scalar charges increase as the separation decreases.
For \(\mu=0.1\), both components still originate from marginal RN black holes, but their induced responses differ. The smaller black hole develops a larger normalized scalar charge than its more massive companion, as shown by the orange curves.

We next consider binaries consisting of a scalarized F3 black hole and an RN F1 black hole, as defined in Tab.~\ref{tab:binary-families} and shown in Figs.~\ref{fig:binary-B1}(b) and~\ref{fig:binary-B1}(c). The F3 black hole already carries scalar charge at zero source, and its field induces a scalar charge on the RN companion. At large separation, the induced charge is approximately
\begin{equation}
Q_{s}^{\mathrm{F1}}\simeq
\frac{\chi_s^{\mathrm{F1}}(0)Q_{s}^{\mathrm{F3}}(0)}{d}.
\label{eq:rn-companion-induction}
\end{equation}
The scalar field produced by F1 and acting back on F3 is suppressed by another factor of \(1/d\). The mutual feedback is therefore weak at large separation, and the scalarized black hole remains close to its zero-source state.

As the separation decreases, the stronger companion field initially increases the scalar charge of the RN black hole. The charge then reaches a maximum and decreases, a behavior particularly evident for the smaller RN companion at \(\mu=0.1\) in Figs.~\ref{fig:binary-B1}(b) and~\ref{fig:binary-B1}(c). Beyond the maximum, the isolated response has \(\chi_s<0\), so a stronger external field produces a smaller scalar charge. The scalar field at the RN horizon, however, continues to increase. This is because the horizon value includes the ambient field supplied by the companion, whereas \(Q_s\) measures the coefficient of the asymptotic \(1/r\) term relative to that background. The scalar field near the horizon can therefore grow even as the scalar charge decreases.

\begin{figure*}[t]
\centering
\includegraphics[width=\textwidth]{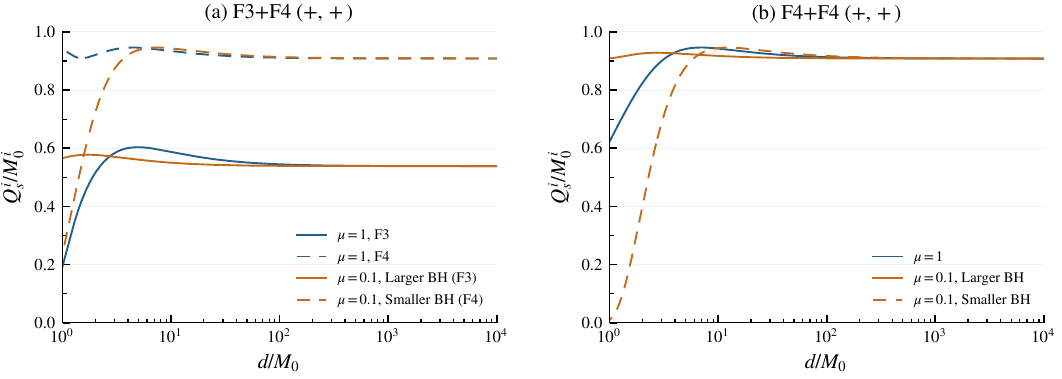}

\caption{Aligned \((+,+)\) equilibrium scalar responses of (a) F3+F4 and (b) F4+F4 binaries, with both black holes carrying \(Q_s>0\) at zero source. The normalized scalar charges \(Q_{s}^{i}/M_{0}^{i}\) are shown as functions of the separation \(d/M_0\). Blue and orange curves correspond to \(\mu=1\) and \(0.1\), respectively. In (a), solid curves represent F3 (A) and dashed curves F4 (B), with F3 being the larger black hole at \(\mu=0.1\). In (b), the two identical black holes at \(\mu=1\) have the same scalar charge and are represented by a single blue curve. At \(\mu=0.1\), solid and dashed orange curves represent the larger and smaller F4 black holes, respectively. In both pairings, the smaller black hole exhibits a stronger variation in its scalar charge as the separation decreases, while the larger black hole remains closer to its zero-source state. Here \(M_{0}^{i}\) is the zero-source mass of each component and \(M_0=M_0^A+M_0^B\).}
\label{fig:binary-B2}
\end{figure*}

We finally consider binaries in which both black holes are already scalarized at zero source. Their scalar charges may have the same or opposite signs, corresponding to aligned and anti-aligned configurations, respectively. We begin with the aligned \((+,+)\) binaries in Fig.~\ref{fig:binary-B2}, whose components both have \(Q_s>0\) at zero source. These binaries pair F3 with F4 or two F4 black holes from Tab.~\ref{tab:binary-families}. At large separation, the positive scalar field of each companion enhances the other's scalar charge. As the separation decreases, however, this field may push a black hole beyond the maximum of its isolated scalar response, causing its charge to decrease even as the external field grows.

In the F3+F4 binaries of Fig.~\ref{fig:binary-B2}(a), F4 carries a larger scalar charge per unit mass than F3 at zero source. For \(\mu=1\), the two black holes have equal masses, so F4 also carries the larger total scalar charge and produces the stronger field acting on F3. Consequently, F3 undergoes the greater change in scalar charge, with a substantial decrease at small separation, while F4 changes comparatively little. For \(\mu=0.1\), F3 is ten times more massive than F4 at zero source, reversing the ordering of their total scalar charges. F3 now generates the stronger companion field and drives a pronounced rise and fall in the charge of the smaller F4 black hole, whereas its own charge remains close to the zero-source value.

In the F4+F4 binaries of Fig.~\ref{fig:binary-B2}(b), both components have the same isolated scalar response. For \(\mu=1\), the two identical black holes carry equal scalar charges along the equilibrium sequence shown, so a single blue curve represents both components. For \(\mu=0.1\), the larger black hole carries the greater total scalar charge and produces a stronger external field at its smaller companion. The smaller black hole consequently exhibits a pronounced rise and fall in its scalar charge as the separation decreases, while the larger black hole undergoes only a modest change. Thus, unequal masses can produce markedly different scalar responses even when both black holes originate from the same isolated family.

\begin{figure*}[t]
\centering
\includegraphics[width=\textwidth]{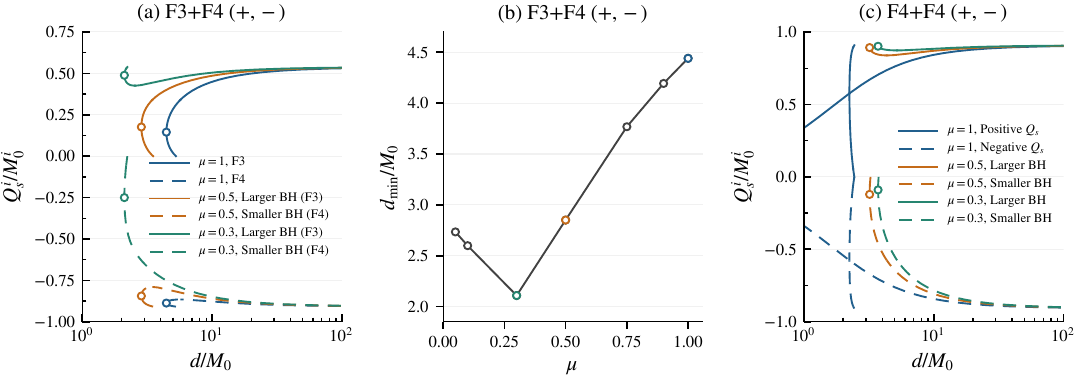}

\caption{Anti-aligned $(+,-)$ binary equilibrium responses. The notation $(+,-)$ indicates that black holes A and B carry positive and negative scalar charges, respectively, at large separation. The scalar charges are normalized by the corresponding zero-source masses, $M_{0}^{i}$.
(a) Scalar charges for F3+F4 binaries at $\mu=1$ (blue), $0.5$ (orange), and $0.3$ (green). Solid curves correspond to A (F3), and dashed curves to B (F4).
(b) Minimum equilibrium separation $d_{\min}/M_0$ for F3+F4 binaries as a function of the mass ratio.
(c) Scalar charges for F4+F4 binaries at the same mass ratios, with solid and dashed curves denoting A and B, respectively. For $\mu=1$, the blue curves show a sequence with equal scalar-charge
magnitudes and two asymmetric branches with their
scalar-charge magnitudes interchanged. For $\mu=0.5$ and $0.3$, both components exhibit returning branches, although the larger black hole undergoes only a small change in scalar charge.
In (a) and (c), paired open circles mark the minimum separation of an equilibrium sequence, with one circle for each black hole.}

\label{fig:binary-B3}
\end{figure*}

Fig.~\ref{fig:binary-B3} shows anti-aligned F3+F4 and F4+F4 binaries, in which the two black holes carry scalar charges of opposite signs at large separation. We adopt the convention \(Q_s^A>0\) and \(Q_s^B<0\), with A denoting F3 and B denoting F4 in the mixed-family binaries. The opposite sign assignment is equivalent under the symmetry \(\phi\to-\phi\). Unlike the aligned binaries in Fig.~\ref{fig:binary-B2}, several anti-aligned equilibrium sequences develop a minimum separation \(d_{\min}\), at which the sequence connected to widely separated black holes turns back toward larger separations.

The origin of this behavior can be understood by considering
a binary whose black holes have very different magnitudes
of scalar charge. At a given separation, the black hole
with the smaller magnitude experiences the stronger field
generated by its companion. When the scalar charges have
opposite signs, this field tends to suppress the original
charge. The degree of suppression, however, also depends
on the nonlinear response of each black hole, which may
become particularly important when their scalar charges
are comparable in magnitude. As the separation decreases,
one component may therefore lose much of its scalar
charge while the other remains comparatively unaffected. The isolated solutions in Sec.~\ref{sec:existence} show that the branch retaining its original scalar-charge sign cannot persist under arbitrarily strong external fields of the opposite sign. This provides a physical explanation for why the anti-aligned equilibrium sequence can turn around at a finite separation, even while both black holes retain nonzero scalar hair. %

The F3+F4 binaries in Fig.~\ref{fig:binary-B3}(a)
illustrate how the suppression depends on the mass ratio.
For equal masses, F4 carries the larger magnitude of
scalar charge at large separation and produces a stronger
external field acting on F3. As the separation decreases,
F3 loses much of its scalar charge, whereas F4 retains
most of its original charge. At $\mu=0.5$, F3 still
undergoes greater suppression, even though it already
carries the larger magnitude of total scalar charge at
large separation. This is because the suppression depends
not only on the strength of the companion field but also
on the scalar response of each black hole.
At $\mu=0.3$, F4 instead undergoes greater suppression,
while the more massive F3 black hole retains most of
its original charge. As shown in
Fig.~\ref{fig:binary-B3}(b), the minimum equilibrium
separation also varies nonmonotonically with the mass
ratio, first decreasing and then increasing as
$\mu$ is reduced.

A similar imbalance occurs in unequal-mass F4+F4 binaries, even though both black holes belong to the same zero-source family. As shown in Fig.~\ref{fig:binary-B3}(c), the larger black hole initially carries the greater total scalar-charge magnitude and generates a stronger opposing field at its smaller companion. As the separation decreases, the scalar charge of the smaller black hole is strongly suppressed, while that of the larger black hole remains close to its zero-source value.

The equal-mass F4+F4 binary exhibits a different behavior. Its two black holes have identical zero-source properties but opposite scalar charges, allowing an equilibrium sequence in which their charge magnitudes remain equal. As the separation decreases, both scalar charges are suppressed together. This simultaneous suppression partially compensates for the increase in the companion fields caused by the decreasing separation, allowing the equal-magnitude sequence to continue toward smaller distances without developing a turning point over the range shown. The same binary also admits two asymmetric equilibria
related by exchanging the two black holes and reversing
the sign of the scalar field. In each equilibrium, one
black hole retains a larger magnitude of scalar charge,
while the other experiences greater suppression of its
scalar charge. In one configuration, the black hole with positive scalar charge retains the larger charge magnitude; in the other, the black hole with negative scalar charge does. For the equal-mass case (blue curves) in Fig.~\ref{fig:binary-B3}(c), the upper portions of the solid and dashed curves correspond to one unequal-amplitude equilibrium, while their lower portions correspond to its
symmetry-related counterpart.

\section{Discussion and conclusion}
\label{sec:discussion}

We have studied how external scalar fields affect the existence and scalar charges of static, spherically symmetric black holes in Einstein--Maxwell--scalar theory. Extending the equilibrium response description of Ref.~\cite{Khalil:2019wyy} beyond the vicinity of spontaneous scalarization, we have examined how black holes respond to external fields of different strengths and signs. These isolated responses also provide the basis for studying binaries in which each black hole generates the scalar field acting on its companion.

For black holes with positive scalar charge, a positive external field supports scalar hair below the threshold for spontaneous scalarization, down to arbitrarily small nonzero electric charge. It also allows larger electric charge relative to the mass than in the absence of an external field. A negative external field instead raises the lower existence boundary and lowers the upper one, narrowing the allowed range of electric charge relative to mass. The two boundaries eventually meet, setting a limit on how negative the external field can become while these solutions retain positive scalar charge. Unlike at the RN scalarization threshold, the scalar charge remains nonzero at the lower existence boundary.

The scalar charge does not grow monotonically with an external field of the same sign. Along the equilibrium sequences studied, it initially increases, reaches a maximum, and then decreases, while the susceptibility changes from positive to negative. For an external field of the opposite sign, two distinct black holes can have the same entropy, electric charge, and imposed field but different scalar charges and masses. Their susceptibility diverges as the two solutions approach the point where they meet.

In a binary, two black holes that are RN in isolation can sustain nonzero scalar charges through mutual induction, alongside the solution in which both scalar charges vanish. The dynamical growth of scalar hair in initially RN black hole binaries has also been demonstrated through numerical relativity simulations~\cite{Diaz:2026iyx}. Our equilibrium analysis identifies a critical separation for the onset of mutual induction when the isolated RN black holes have finite scalar susceptibilities. A black hole that is already scalarized can likewise induce hair on an RN companion. When both components carry scalar charges of the same sign, their charges can initially increase and subsequently decrease as the separation becomes smaller.

Binaries with opposite scalar charges exhibit a more intricate equilibrium structure. Several sequences develop a turning point at a finite separation, where the branch connected to widely separated black holes turns back toward larger distances. Two distinct equilibria can then exist at the same separation and merge at the turning point. The location of this turning point depends on the mass ratio and can vary nonmonotonically. For two identical scalarized black holes with opposite charges, however, the sequence with equal charge magnitudes continues toward smaller separations without a turning point over the range studied. The same binary also admits two asymmetric branches,
distinguished by which black hole carries the larger
magnitude of scalar charge. These turning points characterize the scalar equilibrium sequences rather than minimum orbital separations.

Related developments in EsGB gravity provide a useful
perspective on the possible dynamical implications.
A semianalytic approach has successfully predicted
the onset of dynamical scalarization in initially
unscalarized binaries, in agreement with numerical
evolutions~\cite{Capuano:2026lhs}.
For binaries carrying opposite scalar charges,
theoretical studies have shown that the original
scalarized branches may cease to support continued
adiabatic evolution as the separation
decreases~\cite{Julie:2022huo}.
Subsequent numerical simulations have observed
scalar-charge flips during
inspiral~\cite{Lara:2025kzj}.
Our EMS results reveal turning points and multiple
equilibria within the leading-order scalar feedback
model. Whether these equilibria are dynamically stable and can be followed adiabatically during inspiral remains to be determined.
Some of these turning points occur at separations
of only a few times the total mass of the binary,
where the large-separation approximation is no
longer well controlled.
Their quantitative locations and possible dynamical
consequences motivate further studies incorporating
higher-order binary interactions and time-dependent
evolution.

\begin{acknowledgments}
We thank Yiqian Chen and Haitang Yang for helpful
discussions and comments.
This work was supported in part by the National
Natural Science Foundation of China (NSFC) under
Grant Nos.~12275183, 12675075 and 12605116.
Guangzhou Guo acknowledges support from the
China Postdoctoral Science Foundation under
Grant No.~2025M783414.
\end{acknowledgments}

\appendix

\section{Pseudospectral construction of sourced EMS black holes}
\label{app:spectral}

For the ansatz in Eq.~\eqref{eq:ansatz}, we compactify the radial domain and factor out the simple horizon zero of \(N\) by writing
\begin{equation}
x=1-\frac{2r_h}{r},\qquad
N(x)=\frac{1+x}{2}\,n(x).
\label{eq:spectral-coordinate}
\end{equation}
The horizon and spatial infinity correspond to \(x=-1\) and \(x=1\), respectively. The function \(n(x)\) is regular and positive for the nonextremal solutions considered here. Throughout this appendix, primes denote derivatives with respect to \(x\). For convenience, we define
\begin{equation}
\mathcal K(x)=\left(\frac{Q_e}{r_h}\right)^2e^{-\alpha\phi(x)^2}.
\label{eq:spectral-K}
\end{equation}

The radial field equations can be written as
\begin{equation}
\begin{aligned}
0={}&n''+\frac{2n'}{1+x}
-\frac{\mathcal K}{1+x}
\left[1+2\alpha(1-x)\phi\phi'\right]\\
&+(x-1)n'(\phi')^2
-\frac{x+3}{1+x}n(\phi')^2\\
&-2(x-1)^2n(\phi')^4,
\\[1ex]
0={}&\phi''
+\left(\frac{1}{1+x}+\frac{n'}{n}\right)\phi'\\
&+(1-x)(\phi')^3
+\frac{\alpha\mathcal K\phi}{2(1+x)n}.
\end{aligned}
\label{eq:spectral-equations}
\end{equation}
The remaining equations for $\delta$ and $V$ take the first-order form
\begin{equation}
\begin{aligned}
\delta'&=-(1-x)(\phi')^2,\\
V'&=\frac{Q_e}{2r_h}
e^{-\delta-\alpha\phi^2}.
\end{aligned}
\label{eq:spectral-first-integrals}
\end{equation}

At spatial infinity, we impose
\begin{equation}
n(1)=1,\qquad \delta(1)=0,\qquad
V(1)=0,\qquad \phi(1)=\phi_\infty.
\label{eq:spectral-infinity}
\end{equation}
These conditions ensure asymptotic flatness and fix the time normalization, electromagnetic gauge, and external scalar field. At the horizon $x=-1$, regularity gives
\begin{equation}
\begin{aligned}
n'_h-\frac{\mathcal K_h}{2}
+\frac{3\alpha^2\mathcal K_h^2\phi_h^2}{4n_h}&=0,\\
\phi'_h+\frac{\alpha\mathcal K_h\phi_h}{2n_h}&=0,\\
\delta'_h+2(\phi'_h)^2&=0,\\
V'_h-\frac{Q_e}{2r_h}
e^{-\delta_h-\alpha\phi_h^2}&=0.
\end{aligned}
\label{eq:spectral-horizon}
\end{equation}
Here a subscript $h$ denotes evaluation at the horizon. The last two relations are also used as boundary conditions for the differentiated lapse and Maxwell equations in the spectral construction below.

For the pseudospectral construction, we use
Eq.~\eqref{eq:spectral-equations} together with
the equations obtained by differentiating
Eq.~\eqref{eq:spectral-first-integrals}.
The two equations in Eq.~\eqref{eq:spectral-equations}
imply that the residual of the first-order radial
Einstein equation in Eq.~\eqref{eq:mass-radial-equations},
multiplied by \(r\), is constant.
The condition \(n(1)=1\) sets this constant to zero
at spatial infinity, recovering the original
Einstein equation throughout the exterior.
The first-order equations for \(\delta\) and \(V\)
are likewise recovered by imposing their horizon
conditions in Eq.~\eqref{eq:spectral-horizon}.

Each field $u\in\{n,\phi,\delta,V\}$ is represented by a Chebyshev expansion,
\begin{equation}
u(x)=\frac{c_0^{(u)}}{2}
+\sum_{k=1}^{N_c-1}c_k^{(u)}T_k(x),
\label{eq:spectral-expansion}
\end{equation}
where $T_k$ are Chebyshev polynomials and $N_c$ is the number of coefficients per field. The four differential equations are imposed at the $N_c-2$ Chebyshev--Gauss points
\begin{equation}
x_j=-\cos\frac{(2j+1)\pi}{2(N_c-2)},
\qquad j=0,\ldots,N_c-3.
\label{eq:spectral-grid}
\end{equation}
Together with the eight boundary conditions, this gives $4N_c$ equations for the $4N_c$ spectral coefficients.

For sharply varying profiles at high electric charge, we introduce a radial mapping $r/r_h=1+\ell(1+y)/(1-y)$, with $\ell>0$. The Chebyshev expansion and collocation points are then defined in $y$, with derivatives transformed to $x$ by the chain rule. The choice $\ell=1$ recovers $y=x$.

The collocation equations and boundary conditions form a nonlinear system $\boldsymbol R(\boldsymbol c;\phi_\infty)=0$, where $\boldsymbol c$ collects the $4N_c$ spectral coefficients and $(r_h,Q_e,\alpha)$ are held fixed. We solve this system using a damped Newton iteration,
\begin{equation}
J\,\Delta\boldsymbol c=-\boldsymbol R,\qquad
\boldsymbol c_{\rm new}=\boldsymbol c+\gamma\Delta\boldsymbol c.
\label{eq:spectral-newton}
\end{equation}
Here $J=\partial\boldsymbol R/\partial\boldsymbol c$ is the analytic Jacobian, and $0<\gamma\leq1$ is a damping factor chosen to improve convergence. Previously converged solutions are used as initial guesses for neighboring parameter values.

The asymptotic scalar field need not vary monotonically along a family with fixed $(r_h,Q_e,\alpha)$. Near a turning point, we therefore use the horizon scalar $\phi_h$, rather than $\phi_\infty$, as the continuation parameter. Treating $\phi_\infty$ as an additional unknown, we solve the augmented system
\begin{equation}
\boldsymbol R_{\rm aug}
(\boldsymbol c,\phi_\infty;\phi_h)
=
\begin{pmatrix}
\boldsymbol R(\boldsymbol c;\phi_\infty)\\
\phi(-1;\boldsymbol c)-\phi_h
\end{pmatrix}
=0.
\label{eq:spectral-augmented}
\end{equation}
The asymptotic scalar field $\phi_\infty$ is now determined as part of the solution, while the prescribed horizon value $\phi_h$ serves as the continuation parameter. Differentiating the augmented system with respect to $\phi_h$ gives the derivatives of the asymptotic observables along the solution sequence. The scalar susceptibility is then obtained from
\begin{equation}
\chi_s=
\frac{dQ_s/d\phi_h}{d\phi_\infty/d\phi_h}.
\label{eq:spectral-tangent-susceptibility}
\end{equation}
The derivatives are evaluated at fixed $(r_h,Q_e,\alpha)$, corresponding to fixed $(S,Q_e,\alpha)$. This parametrization allows continuation through a turning point where $d\phi_\infty/d\phi_h=0$, provided the augmented system remains regular. At such a point, $\chi_s$ diverges if $dQ_s/d\phi_h\neq0$.

We check the numerical accuracy by increasing the spectral resolution and working precision. We also verify that the solutions satisfy the original field equations at points away from the collocation grid. The scalar susceptibility is checked against finite differences of nearby solutions, while independent shooting calculations provide an additional check of the black hole masses, scalar charges, and field profiles.

\begin{figure*}[t]
\centering
\includegraphics[width=0.9\textwidth]{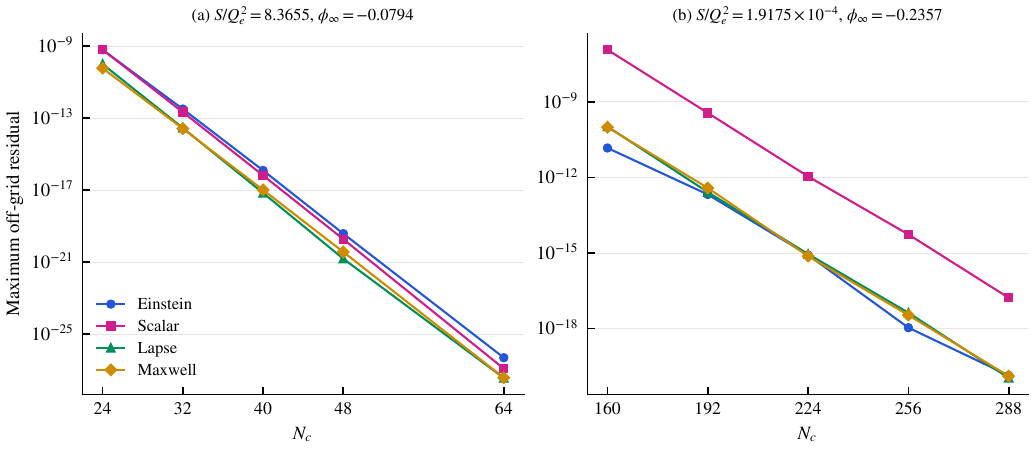}
\caption{Spectral convergence of two sourced EMS black holes, labeled by $S/Q_e^2$ and $\phi_\infty$. Each solution is computed at five spectral resolutions with fixed $(S,Q_e,\alpha,\phi_\infty)$. The maximum residuals of the original field equations are evaluated away from the collocation points. ``Einstein'' denotes the maximum absolute residual among the three Einstein equations, with one index raised and each residual multiplied by $r^2$. ``Scalar'' corresponds to the second equation in Eq.~\eqref{eq:spectral-equations}, while ``Lapse'' and ``Maxwell'' correspond to the first and second equations in Eq.~\eqref{eq:spectral-first-integrals}, respectively. All residuals decrease systematically as $N_c$ increases.}

\label{fig:spectral-convergence}
\end{figure*}

Fig.~\ref{fig:spectral-convergence} presents the spectral convergence for two representative sourced black holes, a moderate-charge background and a high-charge profile. Each solution is computed at five spectral resolutions while keeping $(S,Q_e,\alpha,\phi_\infty)$ fixed. For both backgrounds, the maximum residuals of the original field equations decrease systematically as the number of spectral coefficients $N_c$ increases. The high-charge profile requires substantially higher resolution but exhibits the same systematic convergence. The ADM mass and scalar charge also show consistent convergence.

\section{Physical observables and the mass identity}
\label{app:identities}

Using the compactified coordinate introduced in Appendix~\ref{app:spectral}, the asymptotic expansions $N(r)=1-2M/r+O(r^{-2})$ and $\phi(r)=\phi_\infty+Q_s/r+O(r^{-2})$ give
\begin{equation}
M=r_h\left[\frac{n(1)}{2}+n'(1)\right],\qquad
Q_s=-2r_h\phi'(1).
\label{eq:spectral-asymptotic-observables}
\end{equation}
The Hawking temperature and electric potential difference are
\begin{equation}
T=\frac{n_he^{-\delta_h}}{4\pi r_h},\qquad
\Phi_e=V(1)-V_h=-V_h.
\label{eq:spectral-horizon-observables}
\end{equation}
Here primes denote derivatives with respect to $x$, and a subscript $h$ denotes evaluation at the horizon $x=-1$. We have used the gauge condition $V(1)=0$.

The conserved electric charge can be expressed in terms of the radial Maxwell field as
\begin{equation}
Q_e=2r_he^{\delta+\alpha\phi^2}V'.
\label{eq:spectral-flux}
\end{equation}
The asymptotic electric potential behaves as $V(r)=-Q_{\rm C}/r+O(r^{-2})$, where
\begin{equation}
Q_{\rm C}=2r_hV'(1)=Q_e e^{-\alpha\phi_\infty^2}.
\label{eq:spectral-coulomb}
\end{equation}
Thus, the Coulomb coefficient $Q_{\rm C}$ differs from the conserved charge $Q_e$ when the asymptotic scalar field is nonzero.

We reproduce the finite-source EMS mass identity and the radial first integral from which it follows, as derived in Ref.~\cite{Guo:2026fga}. The identity provides a specific realization of the general scalar-charge relation established in Ref.~\cite{Gibbons:1996af}. To derive the mass identity used in Sec.~\ref{sec:response}, we express the field equations in terms of the areal radius $r$. Using the conserved electric charge $Q_e$ and $f(\phi)=e^{\alpha\phi^2}$, the relevant radial equations are
\begin{equation}
\begin{gathered}
\frac{dN}{dr}
=\frac{1-N}{r}
-rN\left(\frac{d\phi}{dr}\right)^2
-\frac{Q_e^2}{r^3f},\\
\frac{d\delta}{dr}
=-r\left(\frac{d\phi}{dr}\right)^2,\\
\frac{d}{dr}\left(
e^{-\delta}r^2N\frac{d\phi}{dr}
\right)
=-\frac{e^{-\delta}Q_e^2}{2r^2f^2}
\frac{df}{d\phi}.
\end{gathered}
\label{eq:mass-radial-equations}
\end{equation}

The radial equations admit a first integral,
\begin{equation}
\begin{aligned}
\frac{d}{dr}\biggl\{
e^{-2\delta}\biggl[
&r^4\left(\frac12\frac{dN}{dr}
-N\frac{d\delta}{dr}\right)^2\\
&+r^4N^2\left(\frac{d\phi}{dr}\right)^2
-\frac{NQ_e^2}{f(\phi)}
\biggr]\biggr\}=0.
\end{aligned}
\label{eq:mass-first-integral}
\end{equation}
The expression inside the braces is therefore constant throughout the exterior region. Evaluating it at the horizon and at spatial infinity yields an exact relation among the black hole charges and horizon quantities.

At the regular horizon, $N(r_h)=0$, and the conserved quantity reduces to $(2TS)^2$. At spatial infinity, the asymptotic expansions
\begin{equation}
\begin{aligned}
N &= 1-\frac{2M}{r}+O(r^{-2}),\\
\phi &= \phi_\infty+\frac{Q_s}{r}+O(r^{-2}),\\
\delta &= \frac{Q_s^2}{2r^2}+O(r^{-3})
\end{aligned}
\label{eq:mass-asymptotic-expansions}
\end{equation}
give $M^2+Q_s^2-Q_e^2/f(\phi_\infty)$. Equating the two limits yields
\begin{equation}
M^2+Q_s^2
=Q_e^2e^{-\alpha\phi_\infty^2}+(2TS)^2.
\label{eq:sourced-mass-identity}
\end{equation}
Dividing by $Q_e^2$ gives Eq.~\eqref{eq:amplitude-balance} in the main text. At zero source, this reduces to the EMS mass relation of Ref.~\cite{Astefanesei:2019pfq}. The factor $e^{-\alpha\phi_\infty^2}$ reflects the distinction between the conserved electric charge and the asymptotic Coulomb coefficient.

The Maxwell equation gives
\begin{equation}
\Phi_e=\int_{r_h}^{\infty}
\frac{Q_e e^{-\delta}}{r^2f(\phi)}\,dr>0.
\label{eq:app-electric-potential}
\end{equation}
Together with the Smarr relation $M=2TS+\Phi_e Q_e$, this implies
$M^2-(2TS)^2=\Phi_e Q_e(M+2TS)>0$
for the regular nonextremal black holes considered here.
Combining this inequality with the mass identity
in Eq.~\eqref{eq:sourced-mass-identity} yields the upper bound
\begin{equation}
\left|\frac{Q_s}{Q_e}\right|
<e^{-\alpha\phi_\infty^2/2}\leq1,
\label{eq:app-scalar-charge-bound}
\end{equation}
where the last inequality holds for $\alpha>0$.

\section{Zero-mode boundary scaling}
\label{app:zero-mode}

We derive the behavior of the lower existence boundary $q_{\min}(\phi_\infty)$ as $\phi_\infty\to0^-$, near the RN scalarization threshold. At fixed $S$ and $\alpha$, the local scalar response near the RN zero mode allows us to determine the minimum electric charge $Q_e$ at a prescribed negative source. Since the existence boundary is expressed in terms of $q=Q_e/M$, we then account for the corresponding mass variation using the first law and Smarr relation. This yields the two-thirds-power scaling in Eq.~\eqref{eq:boundary-scaling}, with its coefficient $C_{\mathrm{ZM}}$ determined from the local response coefficients.

At fixed $(S,Q_e)$, the local source--response relation
near the RN scalarization threshold is odd in $Q_s$,
owing to the symmetry $\phi\to-\phi$.
Following Ref.~\cite{Khalil:2019wyy}, we expand it as
\begin{equation}
\phi_\infty
=a_1(Q_e)Q_s+a_3(Q_e)Q_s^3+O(Q_s^5).
\label{eq:app-source-response}
\end{equation}
At the RN bifurcation, the linear coefficient vanishes,
$a_1(Q_{e,\mathrm{bif}})=0$.
Varying $Q_e$ at fixed $S$, we define
$\Delta Q_e=Q_e-Q_{e,\mathrm{bif}}$
and expand $a_1(Q_e)=-\kappa\Delta Q_e+\cdots$,
where $\kappa=-da_1/dQ_e$ is evaluated at the bifurcation.
The local relation then becomes
\begin{equation}
\phi_\infty
=-\kappa\Delta Q_e\,Q_s+a_3Q_s^3
+O\!\left((\Delta Q_e)^2Q_s,
\Delta Q_e Q_s^3,Q_s^5\right),
\label{eq:app-cubic-response}
\end{equation}
where $a_3=a_3(Q_{e,\mathrm{bif}})$.
For the $\alpha=10$ solutions considered here,
both $\kappa$ and $a_3$ are positive.
The cubic term captures the leading nonlinear response
near the scalarization threshold, as also encountered
in catastrophe descriptions of
scalarization~\cite{Harada:1998ge}.
The coefficients $\kappa$ and $a_3$ are determined
from the linear RN response and small-amplitude
nonlinear solutions, respectively, as described below.

At fixed $S$ and negative $\phi_\infty$,
we allow $Q_e$ to vary along the branch with $Q_s>0$.
To leading order, Eq.~\eqref{eq:app-cubic-response} gives
\begin{equation}
\Delta Q_e(Q_s)=
\frac{a_3Q_s^2-\phi_\infty/Q_s}{\kappa}.
\label{eq:app-charge-response}
\end{equation}
The first term increases with $Q_s$, whereas the second increases as $Q_s$ decreases. Their sum reaches a minimum at $2a_3Q_s+\phi_\infty/Q_s^2=0$, yielding
\begin{equation}
\kappa\Delta Q_e=3a_3Q_s^2,\qquad
\phi_\infty=-2a_3Q_s^3.
\label{eq:app-fold}
\end{equation}
The second derivative is $6a_3/\kappa>0$, confirming that this point is a local minimum. Eliminating $Q_s$, we obtain
\begin{equation}
\Delta Q_e=
\frac{3a_3^{1/3}}{2^{2/3}\kappa}
(-\phi_\infty)^{2/3}
+O\!\left((-\phi_\infty)^{4/3}\right).
\label{eq:app-charge-displacement}
\end{equation}
The two-thirds exponent follows from the cubic dependence of $\phi_\infty$ on $Q_s$ and the quadratic dependence of $\Delta Q_e$ on $Q_s$ at the minimum.

The minimum electric charge obtained at fixed $(S,\phi_\infty)$ also corresponds to a turning point of the external field along a fixed-$(S,Q_e)$ sequence. Differentiating the cubic relation at fixed $Q_e$ gives
\begin{equation}
\left(\frac{\partial\phi_\infty}{\partial Q_s}\right)_{S,Q_e}
=-\kappa\Delta Q_e+3a_3Q_s^2+\cdots,
\label{eq:app-source-turning}
\end{equation}
which vanishes to leading order by Eq.~\eqref{eq:app-fold}. Since the second derivative is positive for $Q_s>0$, this turning point is a local minimum of $\phi_\infty$.

The existence boundary in Fig.~\ref{fig:1} is expressed in terms of $q=Q_e/M$, rather than $Q_e$. We therefore need to establish that the minimum of $Q_e$ found above also corresponds to a minimum of $q$. Along a sequence with fixed $S$ and $\phi_\infty$, the first law gives $dM=\Phi_e\,dQ_e$. Differentiating $q$ and using the Smarr relation $M=2TS+\Phi_e Q_e$, we obtain
\begin{equation}
\left(\frac{dq}{dQ_s}\right)_{S,\phi_\infty}
=
\frac{2TS}{M^2}
\left(\frac{dQ_e}{dQ_s}\right)_{S,\phi_\infty}.
\label{eq:fixed-source-q}
\end{equation}
Since $T>0$ for the nonextremal black holes considered here, $q$ and $Q_e$ have the same local monotonicity. Their minima therefore occur at the same point, which determines $q_{\min}(\phi_\infty)$ near the RN bifurcation.

To determine the leading shift in $q=Q_e/M$, we must account for the change in mass due to the scalar field. Let $M_{\mathrm{RN}}$ denote the mass of the RN black hole with the same entropy and electric charge. Along a fixed-$(S,Q_e)$ sequence, the first law gives $dM=-Q_s\,d\phi_\infty$. Integrating from the RN solution using the cubic source relation, we obtain
\begin{equation}
M-M_{\mathrm{RN}}
=\frac{\kappa\Delta Q_e}{2}Q_s^2
-\frac{3a_3}{4}Q_s^4+\cdots.
\label{eq:app-mass-shift}
\end{equation}
At the lower existence boundary,
Eq.~\eqref{eq:app-fold} gives
$Q_s^2=\kappa\Delta Q_e/(3a_3)$.
Substituting into Eq.~\eqref{eq:app-mass-shift},
we obtain
$M-M_{\mathrm{RN}}
=\kappa^2(\Delta Q_e)^2/(12a_3)
+O((\Delta Q_e)^3)$.
The scalar contribution to the mass therefore
enters at quadratic order in $\Delta Q_e$,
so the leading shift in $q$ is determined by
the RN relation at fixed entropy.
For an RN black hole with horizon radius $r_h$,
we have $M_{\mathrm{RN}}=(r_h^2+Q_e^2)/(2r_h)$
and $q_{\mathrm{RN}}=2r_hQ_e/(r_h^2+Q_e^2)$.
Expanding $q_{\mathrm{RN}}$ around the bifurcation
and using Eq.~\eqref{eq:app-charge-displacement},
we obtain
\begin{equation}
\begin{aligned}
C_{\mathrm{ZM}}
&=\left.\frac{dq_{\mathrm{RN}}}{dQ_e}\right|_{S,\mathrm{bif}}
\frac{3a_3^{1/3}}{2^{2/3}\kappa}\\
&=\frac{2r_h(r_h^2-Q_{e,\mathrm{bif}}^2)}
{(r_h^2+Q_{e,\mathrm{bif}}^2)^2}
\frac{3a_3^{1/3}}{2^{2/3}\kappa}.
\end{aligned}
\label{eq:app-prefactor}
\end{equation}
The first factor converts the electric-charge displacement into a shift in $q$, while the second is determined by the local nonlinear scalar response.

The coefficients $\kappa$ and $a_3$ are determined from the local scalar response, independently of the existence-boundary data. At fixed $S$, we obtain $\kappa$ from the negative slope of the linear RN source-to-charge ratio $\phi_\infty/Q_s$ with respect to $Q_e$ at the bifurcation. The cubic coefficient $a_3$ is extracted from small-amplitude nonlinear solutions at $Q_e=Q_{e,\mathrm{bif}}$ by extrapolating $\phi_\infty/Q_s^3$ linearly in $Q_s^2$ to zero amplitude, with higher-order extrapolations providing a consistency check. Using $r_h=1$ as the length convention, we obtain $\kappa\simeq6.2801$ and $a_3\simeq20.6250$. Together with $Q_{e,\mathrm{bif}}\simeq0.3685$, corresponding to $q_{\mathrm{bif}}\simeq0.6489$, Eq.~\eqref{eq:app-prefactor} gives $C_{\mathrm{ZM}}\simeq1.1057$.

We compare the predicted scaling with the numerically determined lower existence boundary. The ratio $(q_{\min}-q_{\mathrm{bif}})/(-\phi_\infty)^{2/3}$ approaches $C_{\mathrm{ZM}}$ as $\phi_\infty\to0^-$. At $\phi_\infty=-10^{-4}$, the numerical ratio is $1.1051$, with a relative difference of $5.8\times10^{-4}$ from the predicted coefficient. At $\phi_\infty=-10^{-6}$, the ratio increases to $1.1057$, and the relative difference decreases to $2.7\times10^{-5}$. These results agree with the predicted two-thirds-power scaling and are consistent with the higher-order correction in Eq.~\eqref{eq:boundary-scaling}.

An independent shooting calculation provides a further check of the approach to the RN zero mode. Along the lower existence boundary, the normalized scalar profile $(\phi-\phi_\infty)/(\phi_h-\phi_\infty)$ approaches the RN zero-mode profile normalized to unity at the horizon as $\phi_\infty\to0^-$. This agreement confirms that the limiting scalar configuration approaches the RN zero mode, complementing the numerical verification of the two-thirds-power scaling.

\onecolumngrid
\clearpage
\twocolumngrid
\bibliographystyle{apsrev4-2}
\bibliography{ref}

\begin{thebibliography}{77}%
\makeatletter
\providecommand \@ifxundefined [1]{%
 \@ifx{#1\undefined}
}%
\providecommand \@ifnum [1]{%
 \ifnum #1\expandafter \@firstoftwo
 \else \expandafter \@secondoftwo
 \fi
}%
\providecommand \@ifx [1]{%
 \ifx #1\expandafter \@firstoftwo
 \else \expandafter \@secondoftwo
 \fi
}%
\providecommand \natexlab [1]{#1}%
\providecommand \enquote  [1]{``#1''}%
\providecommand \bibnamefont  [1]{#1}%
\providecommand \bibfnamefont [1]{#1}%
\providecommand \citenamefont [1]{#1}%
\providecommand \href@noop [0]{\@secondoftwo}%
\providecommand \href [0]{\begingroup \@sanitize@url \@href}%
\providecommand \@href[1]{\@@startlink{#1}\@@href}%
\providecommand \@@href[1]{\endgroup#1\@@endlink}%
\providecommand \@sanitize@url [0]{\catcode `\\12\catcode `\$12\catcode
  `\&12\catcode `\#12\catcode `\^12\catcode `\_12\catcode `\%12\relax}%
\providecommand \@@startlink[1]{}%
\providecommand \@@endlink[0]{}%
\providecommand \url  [0]{\begingroup\@sanitize@url \@url }%
\providecommand \@url [1]{\endgroup\@href {#1}{\urlprefix }}%
\providecommand \urlprefix  [0]{URL }%
\providecommand \Eprint [0]{\href }%
\providecommand \doibase [0]{https://doi.org/}%
\providecommand \selectlanguage [0]{\@gobble}%
\providecommand \bibinfo  [0]{\@secondoftwo}%
\providecommand \bibfield  [0]{\@secondoftwo}%
\providecommand \translation [1]{[#1]}%
\providecommand \BibitemOpen [0]{}%
\providecommand \bibitemStop [0]{}%
\providecommand \bibitemNoStop [0]{.\EOS\space}%
\providecommand \EOS [0]{\spacefactor3000\relax}%
\providecommand \BibitemShut  [1]{\csname bibitem#1\endcsname}%
\let\auto@bib@innerbib\@empty
\bibitem [{\citenamefont {Damour}\ and\ \citenamefont
  {Esposito-Farese}(1993)}]{Damour:1993hw}%
  \BibitemOpen
  \bibfield  {author} {\bibinfo {author} {\bibfnamefont {T.}~\bibnamefont
  {Damour}}\ and\ \bibinfo {author} {\bibfnamefont {G.}~\bibnamefont
  {Esposito-Farese}},\ }\href {https://doi.org/10.1103/PhysRevLett.70.2220}
  {\bibfield  {journal} {\bibinfo  {journal} {Phys. Rev. Lett.}\ }\textbf
  {\bibinfo {volume} {70}},\ \bibinfo {pages} {2220} (\bibinfo {year}
  {1993})}\BibitemShut {NoStop}%
\bibitem [{\citenamefont {Damour}\ and\ \citenamefont
  {Esposito-Farese}(1996)}]{Damour:1996ke}%
  \BibitemOpen
  \bibfield  {author} {\bibinfo {author} {\bibfnamefont {T.}~\bibnamefont
  {Damour}}\ and\ \bibinfo {author} {\bibfnamefont {G.}~\bibnamefont
  {Esposito-Farese}},\ }\href {https://doi.org/10.1103/PhysRevD.54.1474}
  {\bibfield  {journal} {\bibinfo  {journal} {Phys. Rev. D}\ }\textbf {\bibinfo
  {volume} {54}},\ \bibinfo {pages} {1474} (\bibinfo {year} {1996})},\ \Eprint
  {https://arxiv.org/abs/gr-qc/9602056} {arXiv:gr-qc/9602056} \BibitemShut
  {NoStop}%
\bibitem [{\citenamefont {Harada}(1998)}]{Harada:1998ge}%
  \BibitemOpen
  \bibfield  {author} {\bibinfo {author} {\bibfnamefont {T.}~\bibnamefont
  {Harada}},\ }\href {https://doi.org/10.1103/PhysRevD.57.4802} {\bibfield
  {journal} {\bibinfo  {journal} {Phys. Rev. D}\ }\textbf {\bibinfo {volume}
  {57}},\ \bibinfo {pages} {4802} (\bibinfo {year} {1998})},\ \Eprint
  {https://arxiv.org/abs/gr-qc/9801049} {arXiv:gr-qc/9801049} \BibitemShut
  {NoStop}%
\bibitem [{\citenamefont {Gubser}(2005)}]{Gubser:2005ih}%
  \BibitemOpen
  \bibfield  {author} {\bibinfo {author} {\bibfnamefont {S.~S.}\ \bibnamefont
  {Gubser}},\ }\href {https://doi.org/10.1088/0264-9381/22/23/013} {\bibfield
  {journal} {\bibinfo  {journal} {Class. Quant. Grav.}\ }\textbf {\bibinfo
  {volume} {22}},\ \bibinfo {pages} {5121} (\bibinfo {year} {2005})},\ \Eprint
  {https://arxiv.org/abs/hep-th/0505189} {arXiv:hep-th/0505189} \BibitemShut
  {NoStop}%
\bibitem [{\citenamefont {Stefanov}\ \emph {et~al.}(2008)\citenamefont
  {Stefanov}, \citenamefont {Yazadjiev},\ and\ \citenamefont
  {Todorov}}]{Stefanov:2007eq}%
  \BibitemOpen
  \bibfield  {author} {\bibinfo {author} {\bibfnamefont {I.~Z.}\ \bibnamefont
  {Stefanov}}, \bibinfo {author} {\bibfnamefont {S.~S.}\ \bibnamefont
  {Yazadjiev}},\ and\ \bibinfo {author} {\bibfnamefont {M.~D.}\ \bibnamefont
  {Todorov}},\ }\href {https://doi.org/10.1142/S0217732308028351} {\bibfield
  {journal} {\bibinfo  {journal} {Mod. Phys. Lett. A}\ }\textbf {\bibinfo
  {volume} {23}},\ \bibinfo {pages} {2915} (\bibinfo {year} {2008})},\ \Eprint
  {https://arxiv.org/abs/0708.4141} {arXiv:0708.4141 [gr-qc]} \BibitemShut
  {NoStop}%
\bibitem [{\citenamefont {Doneva}\ \emph {et~al.}(2010)\citenamefont {Doneva},
  \citenamefont {Yazadjiev}, \citenamefont {Kokkotas},\ and\ \citenamefont
  {Stefanov}}]{Doneva:2010ke}%
  \BibitemOpen
  \bibfield  {author} {\bibinfo {author} {\bibfnamefont {D.~D.}\ \bibnamefont
  {Doneva}}, \bibinfo {author} {\bibfnamefont {S.~S.}\ \bibnamefont
  {Yazadjiev}}, \bibinfo {author} {\bibfnamefont {K.~D.}\ \bibnamefont
  {Kokkotas}},\ and\ \bibinfo {author} {\bibfnamefont {I.~Z.}\ \bibnamefont
  {Stefanov}},\ }\href {https://doi.org/10.1103/PhysRevD.82.064030} {\bibfield
  {journal} {\bibinfo  {journal} {Phys. Rev. D}\ }\textbf {\bibinfo {volume}
  {82}},\ \bibinfo {pages} {064030} (\bibinfo {year} {2010})},\ \Eprint
  {https://arxiv.org/abs/1007.1767} {arXiv:1007.1767 [gr-qc]} \BibitemShut
  {NoStop}%
\bibitem [{\citenamefont {Cardoso}\ \emph
  {et~al.}(2013{\natexlab{a}})\citenamefont {Cardoso}, \citenamefont {Carucci},
  \citenamefont {Pani},\ and\ \citenamefont {Sotiriou}}]{Cardoso:2013fwa}%
  \BibitemOpen
  \bibfield  {author} {\bibinfo {author} {\bibfnamefont {V.}~\bibnamefont
  {Cardoso}}, \bibinfo {author} {\bibfnamefont {I.~P.}\ \bibnamefont
  {Carucci}}, \bibinfo {author} {\bibfnamefont {P.}~\bibnamefont {Pani}},\ and\
  \bibinfo {author} {\bibfnamefont {T.~P.}\ \bibnamefont {Sotiriou}},\ }\href
  {https://doi.org/10.1103/PhysRevLett.111.111101} {\bibfield  {journal}
  {\bibinfo  {journal} {Phys. Rev. Lett.}\ }\textbf {\bibinfo {volume} {111}},\
  \bibinfo {pages} {111101} (\bibinfo {year} {2013}{\natexlab{a}})},\ \Eprint
  {https://arxiv.org/abs/1308.6587} {arXiv:1308.6587 [gr-qc]} \BibitemShut
  {NoStop}%
\bibitem [{\citenamefont {Cardoso}\ \emph
  {et~al.}(2013{\natexlab{b}})\citenamefont {Cardoso}, \citenamefont {Carucci},
  \citenamefont {Pani},\ and\ \citenamefont {Sotiriou}}]{Cardoso:2013opa}%
  \BibitemOpen
  \bibfield  {author} {\bibinfo {author} {\bibfnamefont {V.}~\bibnamefont
  {Cardoso}}, \bibinfo {author} {\bibfnamefont {I.~P.}\ \bibnamefont
  {Carucci}}, \bibinfo {author} {\bibfnamefont {P.}~\bibnamefont {Pani}},\ and\
  \bibinfo {author} {\bibfnamefont {T.~P.}\ \bibnamefont {Sotiriou}},\ }\href
  {https://doi.org/10.1103/PhysRevD.88.044056} {\bibfield  {journal} {\bibinfo
  {journal} {Phys. Rev. D}\ }\textbf {\bibinfo {volume} {88}},\ \bibinfo
  {pages} {044056} (\bibinfo {year} {2013}{\natexlab{b}})},\ \Eprint
  {https://arxiv.org/abs/1305.6936} {arXiv:1305.6936 [gr-qc]} \BibitemShut
  {NoStop}%
\bibitem [{\citenamefont {Doneva}\ \emph {et~al.}(2024)\citenamefont {Doneva},
  \citenamefont {Ramazano\u{g}lu}, \citenamefont {Silva}, \citenamefont
  {Sotiriou},\ and\ \citenamefont {Yazadjiev}}]{Doneva:2022ewd}%
  \BibitemOpen
  \bibfield  {author} {\bibinfo {author} {\bibfnamefont {D.~D.}\ \bibnamefont
  {Doneva}}, \bibinfo {author} {\bibfnamefont {F.~M.}\ \bibnamefont
  {Ramazano\u{g}lu}}, \bibinfo {author} {\bibfnamefont {H.~O.}\ \bibnamefont
  {Silva}}, \bibinfo {author} {\bibfnamefont {T.~P.}\ \bibnamefont
  {Sotiriou}},\ and\ \bibinfo {author} {\bibfnamefont {S.~S.}\ \bibnamefont
  {Yazadjiev}},\ }\href {https://doi.org/10.1103/RevModPhys.96.015004}
  {\bibfield  {journal} {\bibinfo  {journal} {Rev. Mod. Phys.}\ }\textbf
  {\bibinfo {volume} {96}},\ \bibinfo {pages} {015004} (\bibinfo {year}
  {2024})},\ \Eprint {https://arxiv.org/abs/2211.01766} {arXiv:2211.01766
  [gr-qc]} \BibitemShut {NoStop}%
\bibitem [{\citenamefont {Doneva}\ and\ \citenamefont
  {Yazadjiev}(2018)}]{Doneva:2017bvd}%
  \BibitemOpen
  \bibfield  {author} {\bibinfo {author} {\bibfnamefont {D.~D.}\ \bibnamefont
  {Doneva}}\ and\ \bibinfo {author} {\bibfnamefont {S.~S.}\ \bibnamefont
  {Yazadjiev}},\ }\href {https://doi.org/10.1103/PhysRevLett.120.131103}
  {\bibfield  {journal} {\bibinfo  {journal} {Phys. Rev. Lett.}\ }\textbf
  {\bibinfo {volume} {120}},\ \bibinfo {pages} {131103} (\bibinfo {year}
  {2018})},\ \Eprint {https://arxiv.org/abs/1711.01187} {arXiv:1711.01187
  [gr-qc]} \BibitemShut {NoStop}%
\bibitem [{\citenamefont {Silva}\ \emph {et~al.}(2018)\citenamefont {Silva},
  \citenamefont {Sakstein}, \citenamefont {Gualtieri}, \citenamefont
  {Sotiriou},\ and\ \citenamefont {Berti}}]{Silva:2017uqg}%
  \BibitemOpen
  \bibfield  {author} {\bibinfo {author} {\bibfnamefont {H.~O.}\ \bibnamefont
  {Silva}}, \bibinfo {author} {\bibfnamefont {J.}~\bibnamefont {Sakstein}},
  \bibinfo {author} {\bibfnamefont {L.}~\bibnamefont {Gualtieri}}, \bibinfo
  {author} {\bibfnamefont {T.~P.}\ \bibnamefont {Sotiriou}},\ and\ \bibinfo
  {author} {\bibfnamefont {E.}~\bibnamefont {Berti}},\ }\href
  {https://doi.org/10.1103/PhysRevLett.120.131104} {\bibfield  {journal}
  {\bibinfo  {journal} {Phys. Rev. Lett.}\ }\textbf {\bibinfo {volume} {120}},\
  \bibinfo {pages} {131104} (\bibinfo {year} {2018})},\ \Eprint
  {https://arxiv.org/abs/1711.02080} {arXiv:1711.02080 [gr-qc]} \BibitemShut
  {NoStop}%
\bibitem [{\citenamefont {Antoniou}\ \emph
  {et~al.}(2018{\natexlab{a}})\citenamefont {Antoniou}, \citenamefont
  {Bakopoulos},\ and\ \citenamefont {Kanti}}]{Antoniou:2017acq}%
  \BibitemOpen
  \bibfield  {author} {\bibinfo {author} {\bibfnamefont {G.}~\bibnamefont
  {Antoniou}}, \bibinfo {author} {\bibfnamefont {A.}~\bibnamefont
  {Bakopoulos}},\ and\ \bibinfo {author} {\bibfnamefont {P.}~\bibnamefont
  {Kanti}},\ }\href {https://doi.org/10.1103/PhysRevLett.120.131102} {\bibfield
   {journal} {\bibinfo  {journal} {Phys. Rev. Lett.}\ }\textbf {\bibinfo
  {volume} {120}},\ \bibinfo {pages} {131102} (\bibinfo {year}
  {2018}{\natexlab{a}})},\ \Eprint {https://arxiv.org/abs/1711.03390}
  {arXiv:1711.03390 [hep-th]} \BibitemShut {NoStop}%
\bibitem [{\citenamefont {Antoniou}\ \emph
  {et~al.}(2018{\natexlab{b}})\citenamefont {Antoniou}, \citenamefont
  {Bakopoulos},\ and\ \citenamefont {Kanti}}]{Antoniou:2017hxj}%
  \BibitemOpen
  \bibfield  {author} {\bibinfo {author} {\bibfnamefont {G.}~\bibnamefont
  {Antoniou}}, \bibinfo {author} {\bibfnamefont {A.}~\bibnamefont
  {Bakopoulos}},\ and\ \bibinfo {author} {\bibfnamefont {P.}~\bibnamefont
  {Kanti}},\ }\href {https://doi.org/10.1103/PhysRevD.97.084037} {\bibfield
  {journal} {\bibinfo  {journal} {Phys. Rev. D}\ }\textbf {\bibinfo {volume}
  {97}},\ \bibinfo {pages} {084037} (\bibinfo {year} {2018}{\natexlab{b}})},\
  \Eprint {https://arxiv.org/abs/1711.07431} {arXiv:1711.07431 [hep-th]}
  \BibitemShut {NoStop}%
\bibitem [{\citenamefont {Doneva}\ \emph {et~al.}(2018)\citenamefont {Doneva},
  \citenamefont {Kiorpelidi}, \citenamefont {Nedkova}, \citenamefont
  {Papantonopoulos},\ and\ \citenamefont {Yazadjiev}}]{Doneva:2018rou}%
  \BibitemOpen
  \bibfield  {author} {\bibinfo {author} {\bibfnamefont {D.~D.}\ \bibnamefont
  {Doneva}}, \bibinfo {author} {\bibfnamefont {S.}~\bibnamefont {Kiorpelidi}},
  \bibinfo {author} {\bibfnamefont {P.~G.}\ \bibnamefont {Nedkova}}, \bibinfo
  {author} {\bibfnamefont {E.}~\bibnamefont {Papantonopoulos}},\ and\ \bibinfo
  {author} {\bibfnamefont {S.~S.}\ \bibnamefont {Yazadjiev}},\ }\href
  {https://doi.org/10.1103/PhysRevD.98.104056} {\bibfield  {journal} {\bibinfo
  {journal} {Phys. Rev. D}\ }\textbf {\bibinfo {volume} {98}},\ \bibinfo
  {pages} {104056} (\bibinfo {year} {2018})},\ \Eprint
  {https://arxiv.org/abs/1809.00844} {arXiv:1809.00844 [gr-qc]} \BibitemShut
  {NoStop}%
\bibitem [{\citenamefont {Cunha}\ \emph {et~al.}(2019)\citenamefont {Cunha},
  \citenamefont {Herdeiro},\ and\ \citenamefont {Radu}}]{Cunha:2019dwb}%
  \BibitemOpen
  \bibfield  {author} {\bibinfo {author} {\bibfnamefont {P.~V.}\ \bibnamefont
  {Cunha}}, \bibinfo {author} {\bibfnamefont {C.~A.}\ \bibnamefont
  {Herdeiro}},\ and\ \bibinfo {author} {\bibfnamefont {E.}~\bibnamefont
  {Radu}},\ }\href {https://doi.org/10.1103/PhysRevLett.123.011101} {\bibfield
  {journal} {\bibinfo  {journal} {Phys. Rev. Lett.}\ }\textbf {\bibinfo
  {volume} {123}},\ \bibinfo {pages} {011101} (\bibinfo {year} {2019})},\
  \Eprint {https://arxiv.org/abs/1904.09997} {arXiv:1904.09997 [gr-qc]}
  \BibitemShut {NoStop}%
\bibitem [{\citenamefont {Dima}\ \emph {et~al.}(2020)\citenamefont {Dima},
  \citenamefont {Barausse}, \citenamefont {Franchini},\ and\ \citenamefont
  {Sotiriou}}]{Dima:2020yac}%
  \BibitemOpen
  \bibfield  {author} {\bibinfo {author} {\bibfnamefont {A.}~\bibnamefont
  {Dima}}, \bibinfo {author} {\bibfnamefont {E.}~\bibnamefont {Barausse}},
  \bibinfo {author} {\bibfnamefont {N.}~\bibnamefont {Franchini}},\ and\
  \bibinfo {author} {\bibfnamefont {T.~P.}\ \bibnamefont {Sotiriou}},\ }\href
  {https://doi.org/10.1103/PhysRevLett.125.231101} {\bibfield  {journal}
  {\bibinfo  {journal} {Phys. Rev. Lett.}\ }\textbf {\bibinfo {volume} {125}},\
  \bibinfo {pages} {231101} (\bibinfo {year} {2020})},\ \Eprint
  {https://arxiv.org/abs/2006.03095} {arXiv:2006.03095 [gr-qc]} \BibitemShut
  {NoStop}%
\bibitem [{\citenamefont {Berti}\ \emph {et~al.}(2021)\citenamefont {Berti},
  \citenamefont {Collodel}, \citenamefont {Kleihaus},\ and\ \citenamefont
  {Kunz}}]{Berti:2020kgk}%
  \BibitemOpen
  \bibfield  {author} {\bibinfo {author} {\bibfnamefont {E.}~\bibnamefont
  {Berti}}, \bibinfo {author} {\bibfnamefont {L.~G.}\ \bibnamefont {Collodel}},
  \bibinfo {author} {\bibfnamefont {B.}~\bibnamefont {Kleihaus}},\ and\
  \bibinfo {author} {\bibfnamefont {J.}~\bibnamefont {Kunz}},\ }\href
  {https://doi.org/10.1103/PhysRevLett.126.011104} {\bibfield  {journal}
  {\bibinfo  {journal} {Phys. Rev. Lett.}\ }\textbf {\bibinfo {volume} {126}},\
  \bibinfo {pages} {011104} (\bibinfo {year} {2021})},\ \Eprint
  {https://arxiv.org/abs/2009.03905} {arXiv:2009.03905 [gr-qc]} \BibitemShut
  {NoStop}%
\bibitem [{\citenamefont {Doneva}\ and\ \citenamefont
  {Yazadjiev}(2021)}]{Doneva:2021dqn}%
  \BibitemOpen
  \bibfield  {author} {\bibinfo {author} {\bibfnamefont {D.~D.}\ \bibnamefont
  {Doneva}}\ and\ \bibinfo {author} {\bibfnamefont {S.~S.}\ \bibnamefont
  {Yazadjiev}},\ }\href {https://doi.org/10.1103/PhysRevD.103.064024}
  {\bibfield  {journal} {\bibinfo  {journal} {Phys. Rev. D}\ }\textbf {\bibinfo
  {volume} {103}},\ \bibinfo {pages} {064024} (\bibinfo {year} {2021})},\
  \Eprint {https://arxiv.org/abs/2101.03514} {arXiv:2101.03514 [gr-qc]}
  \BibitemShut {NoStop}%
\bibitem [{\citenamefont {East}\ and\ \citenamefont
  {Ripley}(2021)}]{East:2021bqk}%
  \BibitemOpen
  \bibfield  {author} {\bibinfo {author} {\bibfnamefont {W.~E.}\ \bibnamefont
  {East}}\ and\ \bibinfo {author} {\bibfnamefont {J.~L.}\ \bibnamefont
  {Ripley}},\ }\href {https://doi.org/10.1103/PhysRevLett.127.101102}
  {\bibfield  {journal} {\bibinfo  {journal} {Phys. Rev. Lett.}\ }\textbf
  {\bibinfo {volume} {127}},\ \bibinfo {pages} {101102} (\bibinfo {year}
  {2021})},\ \Eprint {https://arxiv.org/abs/2105.08571} {arXiv:2105.08571
  [gr-qc]} \BibitemShut {NoStop}%
\bibitem [{\citenamefont {Herdeiro}\ \emph {et~al.}(2018)\citenamefont
  {Herdeiro}, \citenamefont {Radu}, \citenamefont {Sanchis-Gual},\ and\
  \citenamefont {Font}}]{Herdeiro:2018wub}%
  \BibitemOpen
  \bibfield  {author} {\bibinfo {author} {\bibfnamefont {C.~A.}\ \bibnamefont
  {Herdeiro}}, \bibinfo {author} {\bibfnamefont {E.}~\bibnamefont {Radu}},
  \bibinfo {author} {\bibfnamefont {N.}~\bibnamefont {Sanchis-Gual}},\ and\
  \bibinfo {author} {\bibfnamefont {J.~A.}\ \bibnamefont {Font}},\ }\href
  {https://doi.org/10.1103/PhysRevLett.121.101102} {\bibfield  {journal}
  {\bibinfo  {journal} {Phys. Rev. Lett.}\ }\textbf {\bibinfo {volume} {121}},\
  \bibinfo {pages} {101102} (\bibinfo {year} {2018})},\ \Eprint
  {https://arxiv.org/abs/1806.05190} {arXiv:1806.05190 [gr-qc]} \BibitemShut
  {NoStop}%
\bibitem [{\citenamefont {Bl\'azquez-Salcedo}\ \emph
  {et~al.}(2018)\citenamefont {Bl\'azquez-Salcedo}, \citenamefont {Doneva},
  \citenamefont {Kunz},\ and\ \citenamefont
  {Yazadjiev}}]{Blazquez-Salcedo:2018jnn}%
  \BibitemOpen
  \bibfield  {author} {\bibinfo {author} {\bibfnamefont {J.~L.}\ \bibnamefont
  {Bl\'azquez-Salcedo}}, \bibinfo {author} {\bibfnamefont {D.~D.}\ \bibnamefont
  {Doneva}}, \bibinfo {author} {\bibfnamefont {J.}~\bibnamefont {Kunz}},\ and\
  \bibinfo {author} {\bibfnamefont {S.~S.}\ \bibnamefont {Yazadjiev}},\ }\href
  {https://doi.org/10.1103/PhysRevD.98.084011} {\bibfield  {journal} {\bibinfo
  {journal} {Phys. Rev. D}\ }\textbf {\bibinfo {volume} {98}},\ \bibinfo
  {pages} {084011} (\bibinfo {year} {2018})},\ \Eprint
  {https://arxiv.org/abs/1805.05755} {arXiv:1805.05755 [gr-qc]} \BibitemShut
  {NoStop}%
\bibitem [{\citenamefont {Silva}\ \emph {et~al.}(2019)\citenamefont {Silva},
  \citenamefont {Macedo}, \citenamefont {Sotiriou}, \citenamefont {Gualtieri},
  \citenamefont {Sakstein},\ and\ \citenamefont {Berti}}]{Silva:2018qhn}%
  \BibitemOpen
  \bibfield  {author} {\bibinfo {author} {\bibfnamefont {H.~O.}\ \bibnamefont
  {Silva}}, \bibinfo {author} {\bibfnamefont {C.~F.~B.}\ \bibnamefont
  {Macedo}}, \bibinfo {author} {\bibfnamefont {T.~P.}\ \bibnamefont
  {Sotiriou}}, \bibinfo {author} {\bibfnamefont {L.}~\bibnamefont {Gualtieri}},
  \bibinfo {author} {\bibfnamefont {J.}~\bibnamefont {Sakstein}},\ and\
  \bibinfo {author} {\bibfnamefont {E.}~\bibnamefont {Berti}},\ }\href
  {https://doi.org/10.1103/PhysRevD.99.064011} {\bibfield  {journal} {\bibinfo
  {journal} {Phys. Rev. D}\ }\textbf {\bibinfo {volume} {99}},\ \bibinfo
  {pages} {064011} (\bibinfo {year} {2019})},\ \Eprint
  {https://arxiv.org/abs/1812.05590} {arXiv:1812.05590 [gr-qc]} \BibitemShut
  {NoStop}%
\bibitem [{\citenamefont {Myung}\ and\ \citenamefont
  {Zou}(2019{\natexlab{a}})}]{Myung:2018jvi}%
  \BibitemOpen
  \bibfield  {author} {\bibinfo {author} {\bibfnamefont {Y.~S.}\ \bibnamefont
  {Myung}}\ and\ \bibinfo {author} {\bibfnamefont {D.-C.}\ \bibnamefont
  {Zou}},\ }\href {https://doi.org/10.1016/j.physletb.2019.01.046} {\bibfield
  {journal} {\bibinfo  {journal} {Phys. Lett. B}\ }\textbf {\bibinfo {volume}
  {790}},\ \bibinfo {pages} {400} (\bibinfo {year} {2019}{\natexlab{a}})},\
  \Eprint {https://arxiv.org/abs/1812.03604} {arXiv:1812.03604 [gr-qc]}
  \BibitemShut {NoStop}%
\bibitem [{\citenamefont {Fernandes}\ \emph
  {et~al.}(2019{\natexlab{a}})\citenamefont {Fernandes}, \citenamefont
  {Herdeiro}, \citenamefont {Pombo}, \citenamefont {Radu},\ and\ \citenamefont
  {Sanchis-Gual}}]{Fernandes:2019rez}%
  \BibitemOpen
  \bibfield  {author} {\bibinfo {author} {\bibfnamefont {P.~G.~S.}\
  \bibnamefont {Fernandes}}, \bibinfo {author} {\bibfnamefont {C.~A.~R.}\
  \bibnamefont {Herdeiro}}, \bibinfo {author} {\bibfnamefont {A.~M.}\
  \bibnamefont {Pombo}}, \bibinfo {author} {\bibfnamefont {E.}~\bibnamefont
  {Radu}},\ and\ \bibinfo {author} {\bibfnamefont {N.}~\bibnamefont
  {Sanchis-Gual}},\ }\href {https://doi.org/10.1088/1361-6382/ab23a1}
  {\bibfield  {journal} {\bibinfo  {journal} {Class. Quant. Grav.}\ }\textbf
  {\bibinfo {volume} {36}},\ \bibinfo {pages} {134002} (\bibinfo {year}
  {2019}{\natexlab{a}})},\ \bibinfo {note} {[Erratum: Class.Quant.Grav. 37,
  049501 (2020)]},\ \Eprint {https://arxiv.org/abs/1902.05079}
  {arXiv:1902.05079 [gr-qc]} \BibitemShut {NoStop}%
\bibitem [{\citenamefont {Myung}\ and\ \citenamefont
  {Zou}(2019{\natexlab{b}})}]{Myung:2019oua}%
  \BibitemOpen
  \bibfield  {author} {\bibinfo {author} {\bibfnamefont {Y.~S.}\ \bibnamefont
  {Myung}}\ and\ \bibinfo {author} {\bibfnamefont {D.-C.}\ \bibnamefont
  {Zou}},\ }\href {https://doi.org/10.1140/epjc/s10052-019-7176-7} {\bibfield
  {journal} {\bibinfo  {journal} {Eur. Phys. J. C}\ }\textbf {\bibinfo {volume}
  {79}},\ \bibinfo {pages} {641} (\bibinfo {year} {2019}{\natexlab{b}})},\
  \Eprint {https://arxiv.org/abs/1904.09864} {arXiv:1904.09864 [gr-qc]}
  \BibitemShut {NoStop}%
\bibitem [{\citenamefont {Konoplya}\ and\ \citenamefont
  {Zhidenko}(2019)}]{Konoplya:2019goy}%
  \BibitemOpen
  \bibfield  {author} {\bibinfo {author} {\bibfnamefont {R.~A.}\ \bibnamefont
  {Konoplya}}\ and\ \bibinfo {author} {\bibfnamefont {A.}~\bibnamefont
  {Zhidenko}},\ }\href {https://doi.org/10.1103/PhysRevD.100.044015} {\bibfield
   {journal} {\bibinfo  {journal} {Phys. Rev. D}\ }\textbf {\bibinfo {volume}
  {100}},\ \bibinfo {pages} {044015} (\bibinfo {year} {2019})},\ \Eprint
  {https://arxiv.org/abs/1907.05551} {arXiv:1907.05551 [gr-qc]} \BibitemShut
  {NoStop}%
\bibitem [{\citenamefont {Fernandes}\ \emph
  {et~al.}(2019{\natexlab{b}})\citenamefont {Fernandes}, \citenamefont
  {Herdeiro}, \citenamefont {Pombo}, \citenamefont {Radu},\ and\ \citenamefont
  {Sanchis-Gual}}]{Fernandes:2019kmh}%
  \BibitemOpen
  \bibfield  {author} {\bibinfo {author} {\bibfnamefont {P.~G.}\ \bibnamefont
  {Fernandes}}, \bibinfo {author} {\bibfnamefont {C.~A.}\ \bibnamefont
  {Herdeiro}}, \bibinfo {author} {\bibfnamefont {A.~M.}\ \bibnamefont {Pombo}},
  \bibinfo {author} {\bibfnamefont {E.}~\bibnamefont {Radu}},\ and\ \bibinfo
  {author} {\bibfnamefont {N.}~\bibnamefont {Sanchis-Gual}},\ }\href
  {https://doi.org/10.1103/PhysRevD.100.084045} {\bibfield  {journal} {\bibinfo
   {journal} {Phys. Rev. D}\ }\textbf {\bibinfo {volume} {100}},\ \bibinfo
  {pages} {084045} (\bibinfo {year} {2019}{\natexlab{b}})},\ \Eprint
  {https://arxiv.org/abs/1908.00037} {arXiv:1908.00037 [gr-qc]} \BibitemShut
  {NoStop}%
\bibitem [{\citenamefont {Zou}\ and\ \citenamefont
  {Myung}(2019)}]{Zou:2019bpt}%
  \BibitemOpen
  \bibfield  {author} {\bibinfo {author} {\bibfnamefont {D.-C.}\ \bibnamefont
  {Zou}}\ and\ \bibinfo {author} {\bibfnamefont {Y.~S.}\ \bibnamefont
  {Myung}},\ }\href {https://doi.org/10.1103/PhysRevD.100.124055} {\bibfield
  {journal} {\bibinfo  {journal} {Phys. Rev. D}\ }\textbf {\bibinfo {volume}
  {100}},\ \bibinfo {pages} {124055} (\bibinfo {year} {2019})},\ \Eprint
  {https://arxiv.org/abs/1909.11859} {arXiv:1909.11859 [gr-qc]} \BibitemShut
  {NoStop}%
\bibitem [{\citenamefont {Bl\'azquez-Salcedo}\ \emph
  {et~al.}(2020)\citenamefont {Bl\'azquez-Salcedo}, \citenamefont {Herdeiro},
  \citenamefont {Kunz}, \citenamefont {Pombo},\ and\ \citenamefont
  {Radu}}]{Blazquez-Salcedo:2020nhs}%
  \BibitemOpen
  \bibfield  {author} {\bibinfo {author} {\bibfnamefont {J.~L.}\ \bibnamefont
  {Bl\'azquez-Salcedo}}, \bibinfo {author} {\bibfnamefont {C.~A.}\ \bibnamefont
  {Herdeiro}}, \bibinfo {author} {\bibfnamefont {J.}~\bibnamefont {Kunz}},
  \bibinfo {author} {\bibfnamefont {A.~M.}\ \bibnamefont {Pombo}},\ and\
  \bibinfo {author} {\bibfnamefont {E.}~\bibnamefont {Radu}},\ }\href
  {https://doi.org/10.1016/j.physletb.2020.135493} {\bibfield  {journal}
  {\bibinfo  {journal} {Phys. Lett. B}\ }\textbf {\bibinfo {volume} {806}},\
  \bibinfo {pages} {135493} (\bibinfo {year} {2020})},\ \Eprint
  {https://arxiv.org/abs/2002.00963} {arXiv:2002.00963 [gr-qc]} \BibitemShut
  {NoStop}%
\bibitem [{\citenamefont {Fernandes}(2020)}]{Fernandes:2020gay}%
  \BibitemOpen
  \bibfield  {author} {\bibinfo {author} {\bibfnamefont {P.~G.}\ \bibnamefont
  {Fernandes}},\ }\href {https://doi.org/10.1016/j.dark.2020.100716} {\bibfield
   {journal} {\bibinfo  {journal} {Phys. Dark Univ.}\ }\textbf {\bibinfo
  {volume} {30}},\ \bibinfo {pages} {100716} (\bibinfo {year} {2020})},\
  \Eprint {https://arxiv.org/abs/2003.01045} {arXiv:2003.01045 [gr-qc]}
  \BibitemShut {NoStop}%
\bibitem [{\citenamefont {Luis Bl\'azquez-Salcedo}\ \emph
  {et~al.}(2021)\citenamefont {Luis Bl\'azquez-Salcedo}, \citenamefont
  {Herdeiro}, \citenamefont {Kahlen}, \citenamefont {Kunz}, \citenamefont
  {Pombo},\ and\ \citenamefont {Radu}}]{LuisBlazquez-Salcedo:2020rqp}%
  \BibitemOpen
  \bibfield  {author} {\bibinfo {author} {\bibfnamefont {J.}~\bibnamefont {Luis
  Bl\'azquez-Salcedo}}, \bibinfo {author} {\bibfnamefont {C.~A.~R.}\
  \bibnamefont {Herdeiro}}, \bibinfo {author} {\bibfnamefont {S.}~\bibnamefont
  {Kahlen}}, \bibinfo {author} {\bibfnamefont {J.}~\bibnamefont {Kunz}},
  \bibinfo {author} {\bibfnamefont {A.~M.}\ \bibnamefont {Pombo}},\ and\
  \bibinfo {author} {\bibfnamefont {E.}~\bibnamefont {Radu}},\ }\href
  {https://doi.org/10.1140/epjc/s10052-021-08952-w} {\bibfield  {journal}
  {\bibinfo  {journal} {Eur. Phys. J. C}\ }\textbf {\bibinfo {volume} {81}},\
  \bibinfo {pages} {155} (\bibinfo {year} {2021})},\ \Eprint
  {https://arxiv.org/abs/2008.11744} {arXiv:2008.11744 [gr-qc]} \BibitemShut
  {NoStop}%
\bibitem [{\citenamefont {Myung}\ and\ \citenamefont
  {Zou}(2020)}]{Myung:2020dqt}%
  \BibitemOpen
  \bibfield  {author} {\bibinfo {author} {\bibfnamefont {Y.~S.}\ \bibnamefont
  {Myung}}\ and\ \bibinfo {author} {\bibfnamefont {D.-C.}\ \bibnamefont
  {Zou}},\ }\href {https://doi.org/10.1016/j.physletb.2020.135905} {\bibfield
  {journal} {\bibinfo  {journal} {Phys. Lett. B}\ }\textbf {\bibinfo {volume}
  {811}},\ \bibinfo {pages} {135905} (\bibinfo {year} {2020})},\ \Eprint
  {https://arxiv.org/abs/2009.05193} {arXiv:2009.05193 [gr-qc]} \BibitemShut
  {NoStop}%
\bibitem [{\citenamefont {Myung}\ and\ \citenamefont
  {Zou}(2021)}]{Myung:2020ctt}%
  \BibitemOpen
  \bibfield  {author} {\bibinfo {author} {\bibfnamefont {Y.~S.}\ \bibnamefont
  {Myung}}\ and\ \bibinfo {author} {\bibfnamefont {D.-C.}\ \bibnamefont
  {Zou}},\ }\href {https://doi.org/10.1103/PhysRevD.103.024010} {\bibfield
  {journal} {\bibinfo  {journal} {Phys. Rev. D}\ }\textbf {\bibinfo {volume}
  {103}},\ \bibinfo {pages} {024010} (\bibinfo {year} {2021})},\ \Eprint
  {https://arxiv.org/abs/2011.09665} {arXiv:2011.09665 [gr-qc]} \BibitemShut
  {NoStop}%
\bibitem [{\citenamefont {Wang}\ \emph {et~al.}(2021)\citenamefont {Wang},
  \citenamefont {Wu},\ and\ \citenamefont {Yang}}]{Wang:2020ohb}%
  \BibitemOpen
  \bibfield  {author} {\bibinfo {author} {\bibfnamefont {P.}~\bibnamefont
  {Wang}}, \bibinfo {author} {\bibfnamefont {H.}~\bibnamefont {Wu}},\ and\
  \bibinfo {author} {\bibfnamefont {H.}~\bibnamefont {Yang}},\ }\href
  {https://doi.org/10.1103/PhysRevD.103.104012} {\bibfield  {journal} {\bibinfo
   {journal} {Phys. Rev. D}\ }\textbf {\bibinfo {volume} {103}},\ \bibinfo
  {pages} {104012} (\bibinfo {year} {2021})},\ \Eprint
  {https://arxiv.org/abs/2012.01066} {arXiv:2012.01066 [gr-qc]} \BibitemShut
  {NoStop}%
\bibitem [{\citenamefont {Guo}\ \emph {et~al.}(2021)\citenamefont {Guo},
  \citenamefont {Wang}, \citenamefont {Wu},\ and\ \citenamefont
  {Yang}}]{Guo:2021zed}%
  \BibitemOpen
  \bibfield  {author} {\bibinfo {author} {\bibfnamefont {G.}~\bibnamefont
  {Guo}}, \bibinfo {author} {\bibfnamefont {P.}~\bibnamefont {Wang}}, \bibinfo
  {author} {\bibfnamefont {H.}~\bibnamefont {Wu}},\ and\ \bibinfo {author}
  {\bibfnamefont {H.}~\bibnamefont {Yang}},\ }\href
  {https://doi.org/10.1140/epjc/s10052-021-09614-7} {\bibfield  {journal}
  {\bibinfo  {journal} {Eur. Phys. J. C}\ }\textbf {\bibinfo {volume} {81}},\
  \bibinfo {pages} {864} (\bibinfo {year} {2021})},\ \Eprint
  {https://arxiv.org/abs/2102.04015} {arXiv:2102.04015 [gr-qc]} \BibitemShut
  {NoStop}%
\bibitem [{\citenamefont {Gan}\ \emph {et~al.}(2021{\natexlab{a}})\citenamefont
  {Gan}, \citenamefont {Wang}, \citenamefont {Wu},\ and\ \citenamefont
  {Yang}}]{Gan:2021pwu}%
  \BibitemOpen
  \bibfield  {author} {\bibinfo {author} {\bibfnamefont {Q.}~\bibnamefont
  {Gan}}, \bibinfo {author} {\bibfnamefont {P.}~\bibnamefont {Wang}}, \bibinfo
  {author} {\bibfnamefont {H.}~\bibnamefont {Wu}},\ and\ \bibinfo {author}
  {\bibfnamefont {H.}~\bibnamefont {Yang}},\ }\href
  {https://doi.org/10.1103/PhysRevD.104.024003} {\bibfield  {journal} {\bibinfo
   {journal} {Phys. Rev. D}\ }\textbf {\bibinfo {volume} {104}},\ \bibinfo
  {pages} {024003} (\bibinfo {year} {2021}{\natexlab{a}})},\ \Eprint
  {https://arxiv.org/abs/2104.08703} {arXiv:2104.08703 [gr-qc]} \BibitemShut
  {NoStop}%
\bibitem [{\citenamefont {Gan}\ \emph {et~al.}(2021{\natexlab{b}})\citenamefont
  {Gan}, \citenamefont {Wang}, \citenamefont {Wu},\ and\ \citenamefont
  {Yang}}]{Gan:2021xdl}%
  \BibitemOpen
  \bibfield  {author} {\bibinfo {author} {\bibfnamefont {Q.}~\bibnamefont
  {Gan}}, \bibinfo {author} {\bibfnamefont {P.}~\bibnamefont {Wang}}, \bibinfo
  {author} {\bibfnamefont {H.}~\bibnamefont {Wu}},\ and\ \bibinfo {author}
  {\bibfnamefont {H.}~\bibnamefont {Yang}},\ }\href
  {https://doi.org/10.1103/PhysRevD.104.044049} {\bibfield  {journal} {\bibinfo
   {journal} {Phys. Rev. D}\ }\textbf {\bibinfo {volume} {104}},\ \bibinfo
  {pages} {044049} (\bibinfo {year} {2021}{\natexlab{b}})},\ \Eprint
  {https://arxiv.org/abs/2105.11770} {arXiv:2105.11770 [gr-qc]} \BibitemShut
  {NoStop}%
\bibitem [{\citenamefont {Guo}\ \emph {et~al.}(2022{\natexlab{a}})\citenamefont
  {Guo}, \citenamefont {Wang}, \citenamefont {Wu},\ and\ \citenamefont
  {Yang}}]{Guo:2021enm}%
  \BibitemOpen
  \bibfield  {author} {\bibinfo {author} {\bibfnamefont {G.}~\bibnamefont
  {Guo}}, \bibinfo {author} {\bibfnamefont {P.}~\bibnamefont {Wang}}, \bibinfo
  {author} {\bibfnamefont {H.}~\bibnamefont {Wu}},\ and\ \bibinfo {author}
  {\bibfnamefont {H.}~\bibnamefont {Yang}},\ }\href
  {https://doi.org/10.1007/JHEP06(2022)060} {\bibfield  {journal} {\bibinfo
  {journal} {JHEP}\ }\textbf {\bibinfo {volume} {06}},\ \bibinfo {pages}
  {060}},\ \Eprint {https://arxiv.org/abs/2112.14133} {arXiv:2112.14133
  [gr-qc]} \BibitemShut {NoStop}%
\bibitem [{\citenamefont {Guo}\ \emph {et~al.}(2022{\natexlab{b}})\citenamefont
  {Guo}, \citenamefont {Wang}, \citenamefont {Wu},\ and\ \citenamefont
  {Yang}}]{Guo:2022umh}%
  \BibitemOpen
  \bibfield  {author} {\bibinfo {author} {\bibfnamefont {G.}~\bibnamefont
  {Guo}}, \bibinfo {author} {\bibfnamefont {P.}~\bibnamefont {Wang}}, \bibinfo
  {author} {\bibfnamefont {H.}~\bibnamefont {Wu}},\ and\ \bibinfo {author}
  {\bibfnamefont {H.}~\bibnamefont {Yang}},\ }\href
  {https://doi.org/10.1007/JHEP06(2022)073} {\bibfield  {journal} {\bibinfo
  {journal} {JHEP}\ }\textbf {\bibinfo {volume} {06}},\ \bibinfo {pages}
  {073}},\ \Eprint {https://arxiv.org/abs/2204.00982} {arXiv:2204.00982
  [gr-qc]} \BibitemShut {NoStop}%
\bibitem [{\citenamefont {Guo}\ \emph {et~al.}(2022{\natexlab{c}})\citenamefont
  {Guo}, \citenamefont {Jiang}, \citenamefont {Wang},\ and\ \citenamefont
  {Wu}}]{Guo:2022muy}%
  \BibitemOpen
  \bibfield  {author} {\bibinfo {author} {\bibfnamefont {G.}~\bibnamefont
  {Guo}}, \bibinfo {author} {\bibfnamefont {X.}~\bibnamefont {Jiang}}, \bibinfo
  {author} {\bibfnamefont {P.}~\bibnamefont {Wang}},\ and\ \bibinfo {author}
  {\bibfnamefont {H.}~\bibnamefont {Wu}},\ }\href
  {https://doi.org/10.1103/PhysRevD.105.124064} {\bibfield  {journal} {\bibinfo
   {journal} {Phys. Rev. D}\ }\textbf {\bibinfo {volume} {105}},\ \bibinfo
  {pages} {124064} (\bibinfo {year} {2022}{\natexlab{c}})},\ \Eprint
  {https://arxiv.org/abs/2204.13948} {arXiv:2204.13948 [gr-qc]} \BibitemShut
  {NoStop}%
\bibitem [{\citenamefont {Lai}\ \emph {et~al.}(2022{\natexlab{a}})\citenamefont
  {Lai}, \citenamefont {Myung}, \citenamefont {Yue},\ and\ \citenamefont
  {Zou}}]{Lai:2022spn}%
  \BibitemOpen
  \bibfield  {author} {\bibinfo {author} {\bibfnamefont {M.-Y.}\ \bibnamefont
  {Lai}}, \bibinfo {author} {\bibfnamefont {Y.~S.}\ \bibnamefont {Myung}},
  \bibinfo {author} {\bibfnamefont {R.-H.}\ \bibnamefont {Yue}},\ and\ \bibinfo
  {author} {\bibfnamefont {D.-C.}\ \bibnamefont {Zou}},\ }\href
  {https://doi.org/10.1103/PhysRevD.106.044045} {\bibfield  {journal} {\bibinfo
   {journal} {Phys. Rev. D}\ }\textbf {\bibinfo {volume} {106}},\ \bibinfo
  {pages} {044045} (\bibinfo {year} {2022}{\natexlab{a}})},\ \Eprint
  {https://arxiv.org/abs/2206.11587} {arXiv:2206.11587 [gr-qc]} \BibitemShut
  {NoStop}%
\bibitem [{\citenamefont {Hod}(2022)}]{Hod:2022txa}%
  \BibitemOpen
  \bibfield  {author} {\bibinfo {author} {\bibfnamefont {S.}~\bibnamefont
  {Hod}},\ }\href {https://doi.org/10.1007/JHEP08(2022)272} {\bibfield
  {journal} {\bibinfo  {journal} {JHEP}\ }\textbf {\bibinfo {volume} {08}},\
  \bibinfo {pages} {272}},\ \Eprint {https://arxiv.org/abs/2206.12074}
  {arXiv:2206.12074 [gr-qc]} \BibitemShut {NoStop}%
\bibitem [{\citenamefont {Chen}\ \emph {et~al.}(2022)\citenamefont {Chen},
  \citenamefont {Guo}, \citenamefont {Wang}, \citenamefont {Wu},\ and\
  \citenamefont {Yang}}]{Chen:2022qrw}%
  \BibitemOpen
  \bibfield  {author} {\bibinfo {author} {\bibfnamefont {Y.}~\bibnamefont
  {Chen}}, \bibinfo {author} {\bibfnamefont {G.}~\bibnamefont {Guo}}, \bibinfo
  {author} {\bibfnamefont {P.}~\bibnamefont {Wang}}, \bibinfo {author}
  {\bibfnamefont {H.}~\bibnamefont {Wu}},\ and\ \bibinfo {author}
  {\bibfnamefont {H.}~\bibnamefont {Yang}},\ }\href
  {https://doi.org/10.1007/s11433-022-1986-x} {\bibfield  {journal} {\bibinfo
  {journal} {Sci. China Phys. Mech. Astron.}\ }\textbf {\bibinfo {volume}
  {65}},\ \bibinfo {pages} {120412} (\bibinfo {year} {2022})},\ \Eprint
  {https://arxiv.org/abs/2206.13705} {arXiv:2206.13705 [gr-qc]} \BibitemShut
  {NoStop}%
\bibitem [{\citenamefont {Lai}\ \emph {et~al.}(2022{\natexlab{b}})\citenamefont
  {Lai}, \citenamefont {Myung}, \citenamefont {Yue},\ and\ \citenamefont
  {Zou}}]{Lai:2022ppn}%
  \BibitemOpen
  \bibfield  {author} {\bibinfo {author} {\bibfnamefont {M.-Y.}\ \bibnamefont
  {Lai}}, \bibinfo {author} {\bibfnamefont {Y.~S.}\ \bibnamefont {Myung}},
  \bibinfo {author} {\bibfnamefont {R.-H.}\ \bibnamefont {Yue}},\ and\ \bibinfo
  {author} {\bibfnamefont {D.-C.}\ \bibnamefont {Zou}},\ }\href
  {https://doi.org/10.1103/PhysRevD.106.084043} {\bibfield  {journal} {\bibinfo
   {journal} {Phys. Rev. D}\ }\textbf {\bibinfo {volume} {106}},\ \bibinfo
  {pages} {084043} (\bibinfo {year} {2022}{\natexlab{b}})},\ \Eprint
  {https://arxiv.org/abs/2208.11849} {arXiv:2208.11849 [gr-qc]} \BibitemShut
  {NoStop}%
\bibitem [{\citenamefont {Guo}\ \emph {et~al.}(2023{\natexlab{a}})\citenamefont
  {Guo}, \citenamefont {Lu}, \citenamefont {Wang}, \citenamefont {Wu},\ and\
  \citenamefont {Yang}}]{Guo:2022ghl}%
  \BibitemOpen
  \bibfield  {author} {\bibinfo {author} {\bibfnamefont {G.}~\bibnamefont
  {Guo}}, \bibinfo {author} {\bibfnamefont {Y.}~\bibnamefont {Lu}}, \bibinfo
  {author} {\bibfnamefont {P.}~\bibnamefont {Wang}}, \bibinfo {author}
  {\bibfnamefont {H.}~\bibnamefont {Wu}},\ and\ \bibinfo {author}
  {\bibfnamefont {H.}~\bibnamefont {Yang}},\ }\href
  {https://doi.org/10.1103/PhysRevD.107.124037} {\bibfield  {journal} {\bibinfo
   {journal} {Phys. Rev. D}\ }\textbf {\bibinfo {volume} {107}},\ \bibinfo
  {pages} {124037} (\bibinfo {year} {2023}{\natexlab{a}})},\ \Eprint
  {https://arxiv.org/abs/2212.12901} {arXiv:2212.12901 [gr-qc]} \BibitemShut
  {NoStop}%
\bibitem [{\citenamefont {Guo}\ \emph {et~al.}(2023{\natexlab{b}})\citenamefont
  {Guo}, \citenamefont {Wang}, \citenamefont {Wu},\ and\ \citenamefont
  {Yang}}]{Guo:2023ivz}%
  \BibitemOpen
  \bibfield  {author} {\bibinfo {author} {\bibfnamefont {G.}~\bibnamefont
  {Guo}}, \bibinfo {author} {\bibfnamefont {P.}~\bibnamefont {Wang}}, \bibinfo
  {author} {\bibfnamefont {H.}~\bibnamefont {Wu}},\ and\ \bibinfo {author}
  {\bibfnamefont {H.}~\bibnamefont {Yang}},\ }\href
  {https://doi.org/10.1007/JHEP07(2023)070} {\bibfield  {journal} {\bibinfo
  {journal} {JHEP}\ }\textbf {\bibinfo {volume} {07}},\ \bibinfo {pages}
  {070}},\ \Eprint {https://arxiv.org/abs/2301.06483} {arXiv:2301.06483
  [gr-qc]} \BibitemShut {NoStop}%
\bibitem [{\citenamefont {Guo}\ \emph {et~al.}(2023{\natexlab{c}})\citenamefont
  {Guo}, \citenamefont {Wang}, \citenamefont {Wu},\ and\ \citenamefont
  {Yang}}]{Guo:2023mda}%
  \BibitemOpen
  \bibfield  {author} {\bibinfo {author} {\bibfnamefont {G.}~\bibnamefont
  {Guo}}, \bibinfo {author} {\bibfnamefont {P.}~\bibnamefont {Wang}}, \bibinfo
  {author} {\bibfnamefont {H.}~\bibnamefont {Wu}},\ and\ \bibinfo {author}
  {\bibfnamefont {H.}~\bibnamefont {Yang}},\ }\href
  {https://doi.org/10.1007/JHEP10(2023)076} {\bibfield  {journal} {\bibinfo
  {journal} {JHEP}\ }\textbf {\bibinfo {volume} {10}},\ \bibinfo {pages}
  {076}},\ \Eprint {https://arxiv.org/abs/2307.12210} {arXiv:2307.12210
  [gr-qc]} \BibitemShut {NoStop}%
\bibitem [{\citenamefont {Belkhadria}\ and\ \citenamefont
  {Pombo}(2024)}]{Belkhadria:2023ooc}%
  \BibitemOpen
  \bibfield  {author} {\bibinfo {author} {\bibfnamefont {Z.}~\bibnamefont
  {Belkhadria}}\ and\ \bibinfo {author} {\bibfnamefont {A.~M.}\ \bibnamefont
  {Pombo}},\ }\href {https://doi.org/10.1103/PhysRevD.110.044014} {\bibfield
  {journal} {\bibinfo  {journal} {Phys. Rev. D}\ }\textbf {\bibinfo {volume}
  {110}},\ \bibinfo {pages} {044014} (\bibinfo {year} {2024})},\ \Eprint
  {https://arxiv.org/abs/2311.15850} {arXiv:2311.15850 [gr-qc]} \BibitemShut
  {NoStop}%
\bibitem [{\citenamefont {Chen}\ \emph
  {et~al.}(2024{\natexlab{a}})\citenamefont {Chen}, \citenamefont {Wang},\ and\
  \citenamefont {Yang}}]{Chen:2023qic}%
  \BibitemOpen
  \bibfield  {author} {\bibinfo {author} {\bibfnamefont {Y.}~\bibnamefont
  {Chen}}, \bibinfo {author} {\bibfnamefont {P.}~\bibnamefont {Wang}},\ and\
  \bibinfo {author} {\bibfnamefont {H.}~\bibnamefont {Yang}},\ }\href
  {https://doi.org/10.1103/PhysRevD.110.044020} {\bibfield  {journal} {\bibinfo
   {journal} {Phys. Rev. D}\ }\textbf {\bibinfo {volume} {110}},\ \bibinfo
  {pages} {044020} (\bibinfo {year} {2024}{\natexlab{a}})},\ \Eprint
  {https://arxiv.org/abs/2312.10304} {arXiv:2312.10304 [gr-qc]} \BibitemShut
  {NoStop}%
\bibitem [{\citenamefont {Xiong}\ \emph {et~al.}(2024)\citenamefont {Xiong},
  \citenamefont {Zhang},\ and\ \citenamefont {Li}}]{Xiong:2023bpl}%
  \BibitemOpen
  \bibfield  {author} {\bibinfo {author} {\bibfnamefont {W.}~\bibnamefont
  {Xiong}}, \bibinfo {author} {\bibfnamefont {C.-Y.}\ \bibnamefont {Zhang}},\
  and\ \bibinfo {author} {\bibfnamefont {P.-C.}\ \bibnamefont {Li}},\ }\href
  {https://doi.org/10.1088/1475-7516/2024/09/031} {\bibfield  {journal}
  {\bibinfo  {journal} {JCAP}\ }\textbf {\bibinfo {volume} {09}},\ \bibinfo
  {pages} {031}},\ \Eprint {https://arxiv.org/abs/2312.11879} {arXiv:2312.11879
  [gr-qc]} \BibitemShut {NoStop}%
\bibitem [{\citenamefont {Chen}\ \emph
  {et~al.}(2024{\natexlab{b}})\citenamefont {Chen}, \citenamefont {Wang},\ and\
  \citenamefont {Yang}}]{Chen:2024ilc}%
  \BibitemOpen
  \bibfield  {author} {\bibinfo {author} {\bibfnamefont {Y.}~\bibnamefont
  {Chen}}, \bibinfo {author} {\bibfnamefont {P.}~\bibnamefont {Wang}},\ and\
  \bibinfo {author} {\bibfnamefont {H.}~\bibnamefont {Yang}},\ }\href
  {https://doi.org/10.1140/epjc/s10052-024-12635-7} {\bibfield  {journal}
  {\bibinfo  {journal} {Eur. Phys. J. C}\ }\textbf {\bibinfo {volume} {84}},\
  \bibinfo {pages} {270} (\bibinfo {year} {2024}{\natexlab{b}})},\ \Eprint
  {https://arxiv.org/abs/2401.10905} {arXiv:2401.10905 [gr-qc]} \BibitemShut
  {NoStop}%
\bibitem [{\citenamefont {Melis}\ \emph {et~al.}(2025)\citenamefont {Melis},
  \citenamefont {Corelli}, \citenamefont {Croft},\ and\ \citenamefont
  {Pani}}]{Melis:2024kfr}%
  \BibitemOpen
  \bibfield  {author} {\bibinfo {author} {\bibfnamefont {M.}~\bibnamefont
  {Melis}}, \bibinfo {author} {\bibfnamefont {F.}~\bibnamefont {Corelli}},
  \bibinfo {author} {\bibfnamefont {R.}~\bibnamefont {Croft}},\ and\ \bibinfo
  {author} {\bibfnamefont {P.}~\bibnamefont {Pani}},\ }\href
  {https://doi.org/10.1103/PhysRevD.111.064072} {\bibfield  {journal} {\bibinfo
   {journal} {Phys. Rev. D}\ }\textbf {\bibinfo {volume} {111}},\ \bibinfo
  {pages} {064072} (\bibinfo {year} {2025})},\ \Eprint
  {https://arxiv.org/abs/2412.14259} {arXiv:2412.14259 [gr-qc]} \BibitemShut
  {NoStop}%
\bibitem [{\citenamefont {Zhang}\ \emph {et~al.}(2025)\citenamefont {Zhang},
  \citenamefont {Zhang},\ and\ \citenamefont {Zheng}}]{Zhang:2025jlb}%
  \BibitemOpen
  \bibfield  {author} {\bibinfo {author} {\bibfnamefont {C.-Y.}\ \bibnamefont
  {Zhang}}, \bibinfo {author} {\bibfnamefont {Z.}~\bibnamefont {Zhang}},\ and\
  \bibinfo {author} {\bibfnamefont {R.}~\bibnamefont {Zheng}},\ }\href
  {https://doi.org/10.1007/s11433-024-2607-1} {\bibfield  {journal} {\bibinfo
  {journal} {Sci. China Phys. Mech. Astron.}\ }\textbf {\bibinfo {volume}
  {68}},\ \bibinfo {pages} {250411} (\bibinfo {year} {2025})},\ \Eprint
  {https://arxiv.org/abs/2503.08315} {arXiv:2503.08315 [gr-qc]} \BibitemShut
  {NoStop}%
\bibitem [{\citenamefont {Cheng}\ \emph {et~al.}(2025)\citenamefont {Cheng},
  \citenamefont {Guo}, \citenamefont {Wang},\ and\ \citenamefont
  {Yang}}]{Cheng:2025hdw}%
  \BibitemOpen
  \bibfield  {author} {\bibinfo {author} {\bibfnamefont {L.}~\bibnamefont
  {Cheng}}, \bibinfo {author} {\bibfnamefont {G.}~\bibnamefont {Guo}}, \bibinfo
  {author} {\bibfnamefont {P.}~\bibnamefont {Wang}},\ and\ \bibinfo {author}
  {\bibfnamefont {H.}~\bibnamefont {Yang}},\ }\href@noop {} {\  (\bibinfo
  {year} {2025})},\ \Eprint {https://arxiv.org/abs/2506.01773}
  {arXiv:2506.01773 [gr-qc]} \BibitemShut {NoStop}%
\bibitem [{\citenamefont {Wang}\ and\ \citenamefont {Wu}(2026)}]{Wang:2025mxe}%
  \BibitemOpen
  \bibfield  {author} {\bibinfo {author} {\bibfnamefont {P.}~\bibnamefont
  {Wang}}\ and\ \bibinfo {author} {\bibfnamefont {T.}~\bibnamefont {Wu}},\
  }\href {https://doi.org/10.1103/srxw-wsvq} {\bibfield  {journal} {\bibinfo
  {journal} {Phys. Rev. D}\ }\textbf {\bibinfo {volume} {114}},\ \bibinfo
  {pages} {L021501} (\bibinfo {year} {2026})},\ \Eprint
  {https://arxiv.org/abs/2510.12135} {arXiv:2510.12135 [gr-qc]} \BibitemShut
  {NoStop}%
\bibitem [{\citenamefont {Cheng}\ \emph {et~al.}(2026)\citenamefont {Cheng},
  \citenamefont {Guo}, \citenamefont {Li}, \citenamefont {Tao},\ and\
  \citenamefont {Wang}}]{Cheng:2026gxu}%
  \BibitemOpen
  \bibfield  {author} {\bibinfo {author} {\bibfnamefont {L.}~\bibnamefont
  {Cheng}}, \bibinfo {author} {\bibfnamefont {X.}~\bibnamefont {Guo}}, \bibinfo
  {author} {\bibfnamefont {Y.}~\bibnamefont {Li}}, \bibinfo {author}
  {\bibfnamefont {J.}~\bibnamefont {Tao}},\ and\ \bibinfo {author}
  {\bibfnamefont {P.}~\bibnamefont {Wang}},\ }\href@noop {} {\  (\bibinfo
  {year} {2026})},\ \Eprint {https://arxiv.org/abs/2603.22261}
  {arXiv:2603.22261 [gr-qc]} \BibitemShut {NoStop}%
\bibitem [{\citenamefont {Chen}\ \emph {et~al.}(2026)\citenamefont {Chen},
  \citenamefont {Li},\ and\ \citenamefont {Wang}}]{Chen:2026rhr}%
  \BibitemOpen
  \bibfield  {author} {\bibinfo {author} {\bibfnamefont {Y.}~\bibnamefont
  {Chen}}, \bibinfo {author} {\bibfnamefont {L.}~\bibnamefont {Li}},\ and\
  \bibinfo {author} {\bibfnamefont {P.}~\bibnamefont {Wang}},\ }\href
  {https://doi.org/10.1007/s11433-026-3059-4} {\bibfield  {journal} {\bibinfo
  {journal} {Sci. China Phys. Mech. Astron.}\ }\textbf {\bibinfo {volume}
  {69}},\ \bibinfo {pages} {119513} (\bibinfo {year} {2026})},\ \Eprint
  {https://arxiv.org/abs/2604.11652} {arXiv:2604.11652 [gr-qc]} \BibitemShut
  {NoStop}%
\bibitem [{\citenamefont {Khalil}\ \emph {et~al.}(2019)\citenamefont {Khalil},
  \citenamefont {Sennett}, \citenamefont {Steinhoff},\ and\ \citenamefont
  {Buonanno}}]{Khalil:2019wyy}%
  \BibitemOpen
  \bibfield  {author} {\bibinfo {author} {\bibfnamefont {M.}~\bibnamefont
  {Khalil}}, \bibinfo {author} {\bibfnamefont {N.}~\bibnamefont {Sennett}},
  \bibinfo {author} {\bibfnamefont {J.}~\bibnamefont {Steinhoff}},\ and\
  \bibinfo {author} {\bibfnamefont {A.}~\bibnamefont {Buonanno}},\ }\href
  {https://doi.org/10.1103/PhysRevD.100.124013} {\bibfield  {journal} {\bibinfo
   {journal} {Phys. Rev. D}\ }\textbf {\bibinfo {volume} {100}},\ \bibinfo
  {pages} {124013} (\bibinfo {year} {2019})},\ \Eprint
  {https://arxiv.org/abs/1906.08161} {arXiv:1906.08161 [gr-qc]} \BibitemShut
  {NoStop}%
\bibitem [{\citenamefont {Juli{\'e}}\ \emph {et~al.}(2022)\citenamefont
  {Juli{\'e}}, \citenamefont {Silva}, \citenamefont {Berti},\ and\
  \citenamefont {Yunes}}]{Julie:2022huo}%
  \BibitemOpen
  \bibfield  {author} {\bibinfo {author} {\bibfnamefont {F.-L.}\ \bibnamefont
  {Juli{\'e}}}, \bibinfo {author} {\bibfnamefont {H.~O.}\ \bibnamefont
  {Silva}}, \bibinfo {author} {\bibfnamefont {E.}~\bibnamefont {Berti}},\ and\
  \bibinfo {author} {\bibfnamefont {N.}~\bibnamefont {Yunes}},\ }\href
  {https://doi.org/10.1103/PhysRevD.105.124031} {\bibfield  {journal} {\bibinfo
   {journal} {Phys. Rev. D}\ }\textbf {\bibinfo {volume} {105}},\ \bibinfo
  {pages} {124031} (\bibinfo {year} {2022})},\ \Eprint
  {https://arxiv.org/abs/2202.01329} {arXiv:2202.01329 [gr-qc]} \BibitemShut
  {NoStop}%
\bibitem [{\citenamefont {Gibbons}\ \emph {et~al.}(1996)\citenamefont
  {Gibbons}, \citenamefont {Kallosh},\ and\ \citenamefont
  {Kol}}]{Gibbons:1996af}%
  \BibitemOpen
  \bibfield  {author} {\bibinfo {author} {\bibfnamefont {G.~W.}\ \bibnamefont
  {Gibbons}}, \bibinfo {author} {\bibfnamefont {R.}~\bibnamefont {Kallosh}},\
  and\ \bibinfo {author} {\bibfnamefont {B.}~\bibnamefont {Kol}},\ }\href
  {https://doi.org/10.1103/PhysRevLett.77.4992} {\bibfield  {journal} {\bibinfo
   {journal} {Phys. Rev. Lett.}\ }\textbf {\bibinfo {volume} {77}},\ \bibinfo
  {pages} {4992} (\bibinfo {year} {1996})},\ \Eprint
  {https://arxiv.org/abs/hep-th/9607108} {arXiv:hep-th/9607108} \BibitemShut
  {NoStop}%
\bibitem [{\citenamefont {Juli{\'e}}(2023)}]{Julie:2023ncq}%
  \BibitemOpen
  \bibfield  {author} {\bibinfo {author} {\bibfnamefont {F.-L.}\ \bibnamefont
  {Juli{\'e}}},\ }\href@noop {} {\  (\bibinfo {year} {2023})},\ \Eprint
  {https://arxiv.org/abs/2312.16764} {arXiv:2312.16764 [gr-qc]} \BibitemShut
  {NoStop}%
\bibitem [{\citenamefont {Capuano}\ \emph {et~al.}(2026)\citenamefont
  {Capuano}, \citenamefont {Arest{\'e}~Sal{\'o}}, \citenamefont {Doneva},
  \citenamefont {Yazadjiev},\ and\ \citenamefont {Barausse}}]{Capuano:2026lhs}%
  \BibitemOpen
  \bibfield  {author} {\bibinfo {author} {\bibfnamefont {L.}~\bibnamefont
  {Capuano}}, \bibinfo {author} {\bibfnamefont {L.}~\bibnamefont
  {Arest{\'e}~Sal{\'o}}}, \bibinfo {author} {\bibfnamefont {D.~D.}\
  \bibnamefont {Doneva}}, \bibinfo {author} {\bibfnamefont {S.~S.}\
  \bibnamefont {Yazadjiev}},\ and\ \bibinfo {author} {\bibfnamefont
  {E.}~\bibnamefont {Barausse}},\ }\href {https://doi.org/10.1103/hgfh-rwrw}
  {\bibfield  {journal} {\bibinfo  {journal} {Phys. Rev. D}\ }\textbf {\bibinfo
  {volume} {113}},\ \bibinfo {pages} {124038} (\bibinfo {year} {2026})},\
  \Eprint {https://arxiv.org/abs/2602.02650} {arXiv:2602.02650 [gr-qc]}
  \BibitemShut {NoStop}%
\bibitem [{\citenamefont {Silva}\ \emph {et~al.}(2021)\citenamefont {Silva},
  \citenamefont {Witek}, \citenamefont {Elley},\ and\ \citenamefont
  {Yunes}}]{Silva:2020omi}%
  \BibitemOpen
  \bibfield  {author} {\bibinfo {author} {\bibfnamefont {H.~O.}\ \bibnamefont
  {Silva}}, \bibinfo {author} {\bibfnamefont {H.}~\bibnamefont {Witek}},
  \bibinfo {author} {\bibfnamefont {M.}~\bibnamefont {Elley}},\ and\ \bibinfo
  {author} {\bibfnamefont {N.}~\bibnamefont {Yunes}},\ }\href
  {https://doi.org/10.1103/PhysRevLett.127.031101} {\bibfield  {journal}
  {\bibinfo  {journal} {Phys. Rev. Lett.}\ }\textbf {\bibinfo {volume} {127}},\
  \bibinfo {pages} {031101} (\bibinfo {year} {2021})},\ \Eprint
  {https://arxiv.org/abs/2012.10436} {arXiv:2012.10436 [gr-qc]} \BibitemShut
  {NoStop}%
\bibitem [{\citenamefont {Lara}\ \emph {et~al.}(2025)\citenamefont {Lara} \emph
  {et~al.}}]{Lara:2025kzj}%
  \BibitemOpen
  \bibfield  {author} {\bibinfo {author} {\bibfnamefont {G.}~\bibnamefont
  {Lara}} \emph {et~al.},\ }\href@noop {} {\  (\bibinfo {year} {2025})},\
  \Eprint {https://arxiv.org/abs/2505.14785} {arXiv:2505.14785 [gr-qc]}
  \BibitemShut {NoStop}%
\bibitem [{\citenamefont {Guo}\ \emph {et~al.}(2025)\citenamefont {Guo},
  \citenamefont {Qian},\ and\ \citenamefont {Wang}}]{Guo:2025xwh}%
  \BibitemOpen
  \bibfield  {author} {\bibinfo {author} {\bibfnamefont {H.}~\bibnamefont
  {Guo}}, \bibinfo {author} {\bibfnamefont {W.-L.}\ \bibnamefont {Qian}},\ and\
  \bibinfo {author} {\bibfnamefont {B.}~\bibnamefont {Wang}},\ }\href
  {https://doi.org/10.1103/861k-193h} {\bibfield  {journal} {\bibinfo
  {journal} {Phys. Rev. D}\ }\textbf {\bibinfo {volume} {112}},\ \bibinfo
  {pages} {084010} (\bibinfo {year} {2025})},\ \Eprint
  {https://arxiv.org/abs/2506.02445} {arXiv:2506.02445 [gr-qc]} \BibitemShut
  {NoStop}%
\bibitem [{\citenamefont {Xiong}\ \emph {et~al.}(2022)\citenamefont {Xiong},
  \citenamefont {Liu}, \citenamefont {Niu}, \citenamefont {Zhang},\ and\
  \citenamefont {Wang}}]{Xiong:2022ozw}%
  \BibitemOpen
  \bibfield  {author} {\bibinfo {author} {\bibfnamefont {W.}~\bibnamefont
  {Xiong}}, \bibinfo {author} {\bibfnamefont {P.}~\bibnamefont {Liu}}, \bibinfo
  {author} {\bibfnamefont {C.}~\bibnamefont {Niu}}, \bibinfo {author}
  {\bibfnamefont {C.-Y.}\ \bibnamefont {Zhang}},\ and\ \bibinfo {author}
  {\bibfnamefont {B.}~\bibnamefont {Wang}},\ }\href
  {https://doi.org/10.1088/1674-1137/ac70ad} {\bibfield  {journal} {\bibinfo
  {journal} {Chin. Phys. C}\ }\textbf {\bibinfo {volume} {46}},\ \bibinfo
  {pages} {095103} (\bibinfo {year} {2022})},\ \Eprint
  {https://arxiv.org/abs/2205.07538} {arXiv:2205.07538 [gr-qc]} \BibitemShut
  {NoStop}%
\bibitem [{\citenamefont {Niu}\ \emph {et~al.}(2022)\citenamefont {Niu},
  \citenamefont {Xiong}, \citenamefont {Liu}, \citenamefont {Zhang},\ and\
  \citenamefont {Wang}}]{Niu:2022zlf}%
  \BibitemOpen
  \bibfield  {author} {\bibinfo {author} {\bibfnamefont {C.}~\bibnamefont
  {Niu}}, \bibinfo {author} {\bibfnamefont {W.}~\bibnamefont {Xiong}}, \bibinfo
  {author} {\bibfnamefont {P.}~\bibnamefont {Liu}}, \bibinfo {author}
  {\bibfnamefont {C.-Y.}\ \bibnamefont {Zhang}},\ and\ \bibinfo {author}
  {\bibfnamefont {B.}~\bibnamefont {Wang}},\ }\href@noop {} {\  (\bibinfo
  {year} {2022})},\ \Eprint {https://arxiv.org/abs/2209.12117}
  {arXiv:2209.12117 [gr-qc]} \BibitemShut {NoStop}%
\bibitem [{\citenamefont {Corelli}\ \emph {et~al.}(2021)\citenamefont
  {Corelli}, \citenamefont {Ikeda},\ and\ \citenamefont
  {Pani}}]{Corelli:2021ikv}%
  \BibitemOpen
  \bibfield  {author} {\bibinfo {author} {\bibfnamefont {F.}~\bibnamefont
  {Corelli}}, \bibinfo {author} {\bibfnamefont {T.}~\bibnamefont {Ikeda}},\
  and\ \bibinfo {author} {\bibfnamefont {P.}~\bibnamefont {Pani}},\ }\href
  {https://doi.org/10.1103/PhysRevD.104.084069} {\bibfield  {journal} {\bibinfo
   {journal} {Phys. Rev. D}\ }\textbf {\bibinfo {volume} {104}},\ \bibinfo
  {pages} {084069} (\bibinfo {year} {2021})},\ \Eprint
  {https://arxiv.org/abs/2108.08328} {arXiv:2108.08328 [gr-qc]} \BibitemShut
  {NoStop}%
\bibitem [{\citenamefont {Zhang}\ \emph {et~al.}(2022)\citenamefont {Zhang},
  \citenamefont {Chen}, \citenamefont {Liu}, \citenamefont {Luo}, \citenamefont
  {Tian},\ and\ \citenamefont {Wang}}]{Zhang:2021nnn}%
  \BibitemOpen
  \bibfield  {author} {\bibinfo {author} {\bibfnamefont {C.-Y.}\ \bibnamefont
  {Zhang}}, \bibinfo {author} {\bibfnamefont {Q.}~\bibnamefont {Chen}},
  \bibinfo {author} {\bibfnamefont {Y.}~\bibnamefont {Liu}}, \bibinfo {author}
  {\bibfnamefont {W.-K.}\ \bibnamefont {Luo}}, \bibinfo {author} {\bibfnamefont
  {Y.}~\bibnamefont {Tian}},\ and\ \bibinfo {author} {\bibfnamefont
  {B.}~\bibnamefont {Wang}},\ }\href
  {https://doi.org/10.1103/PhysRevLett.128.161105} {\bibfield  {journal}
  {\bibinfo  {journal} {Phys. Rev. Lett.}\ }\textbf {\bibinfo {volume} {128}},\
  \bibinfo {pages} {161105} (\bibinfo {year} {2022})},\ \Eprint
  {https://arxiv.org/abs/2112.07455} {arXiv:2112.07455 [gr-qc]} \BibitemShut
  {NoStop}%
\bibitem [{\citenamefont {Jiang}\ \emph {et~al.}(2024)\citenamefont {Jiang},
  \citenamefont {Chen}, \citenamefont {Liu}, \citenamefont {Tian},
  \citenamefont {Xiong}, \citenamefont {Zhang},\ and\ \citenamefont
  {Wang}}]{Jiang:2023yyn}%
  \BibitemOpen
  \bibfield  {author} {\bibinfo {author} {\bibfnamefont {J.-Y.}\ \bibnamefont
  {Jiang}}, \bibinfo {author} {\bibfnamefont {Q.}~\bibnamefont {Chen}},
  \bibinfo {author} {\bibfnamefont {Y.}~\bibnamefont {Liu}}, \bibinfo {author}
  {\bibfnamefont {Y.}~\bibnamefont {Tian}}, \bibinfo {author} {\bibfnamefont
  {W.}~\bibnamefont {Xiong}}, \bibinfo {author} {\bibfnamefont {C.-Y.}\
  \bibnamefont {Zhang}},\ and\ \bibinfo {author} {\bibfnamefont
  {B.}~\bibnamefont {Wang}},\ }\href
  {https://doi.org/10.1007/s11433-023-2231-5} {\bibfield  {journal} {\bibinfo
  {journal} {Sci. China Phys. Mech. Astron.}\ }\textbf {\bibinfo {volume}
  {67}},\ \bibinfo {pages} {220411} (\bibinfo {year} {2024})},\ \Eprint
  {https://arxiv.org/abs/2306.10371} {arXiv:2306.10371 [gr-qc]} \BibitemShut
  {NoStop}%
\bibitem [{\citenamefont {D{\'\i}az}\ \emph {et~al.}(2026)\citenamefont
  {D{\'\i}az}, \citenamefont {Gonz{\'a}lez}, \citenamefont {Papantonopoulos},\
  and\ \citenamefont {V{\'a}squez}}]{Diaz:2026iyx}%
  \BibitemOpen
  \bibfield  {author} {\bibinfo {author} {\bibfnamefont {J.~P.}\ \bibnamefont
  {D{\'\i}az}}, \bibinfo {author} {\bibfnamefont {P.~A.}\ \bibnamefont
  {Gonz{\'a}lez}}, \bibinfo {author} {\bibfnamefont {E.}~\bibnamefont
  {Papantonopoulos}},\ and\ \bibinfo {author} {\bibfnamefont {Y.}~\bibnamefont
  {V{\'a}squez}},\ }\href@noop {} {\  (\bibinfo {year} {2026})},\ \Eprint
  {https://arxiv.org/abs/2606.21549} {arXiv:2606.21549 [gr-qc]} \BibitemShut
  {NoStop}%
\bibitem [{\citenamefont {Myung}\ and\ \citenamefont
  {Zou}(2019{\natexlab{c}})}]{Myung:2018vug}%
  \BibitemOpen
  \bibfield  {author} {\bibinfo {author} {\bibfnamefont {Y.~S.}\ \bibnamefont
  {Myung}}\ and\ \bibinfo {author} {\bibfnamefont {D.-C.}\ \bibnamefont
  {Zou}},\ }\href {https://doi.org/10.1140/epjc/s10052-019-6792-6} {\bibfield
  {journal} {\bibinfo  {journal} {Eur. Phys. J. C}\ }\textbf {\bibinfo {volume}
  {79}},\ \bibinfo {pages} {273} (\bibinfo {year} {2019}{\natexlab{c}})},\
  \Eprint {https://arxiv.org/abs/1808.02609} {arXiv:1808.02609 [gr-qc]}
  \BibitemShut {NoStop}%
\bibitem [{\citenamefont {Hod}(2019)}]{Hod:2019ulh}%
  \BibitemOpen
  \bibfield  {author} {\bibinfo {author} {\bibfnamefont {S.}~\bibnamefont
  {Hod}},\ }\href {https://doi.org/10.1016/j.physletb.2019.135025} {\bibfield
  {journal} {\bibinfo  {journal} {Phys. Lett. B}\ }\textbf {\bibinfo {volume}
  {798}},\ \bibinfo {pages} {135025} (\bibinfo {year} {2019})},\ \Eprint
  {https://arxiv.org/abs/2002.01948} {arXiv:2002.01948 [gr-qc]} \BibitemShut
  {NoStop}%
\bibitem [{\citenamefont {C{\'a}rdenas}\ \emph {et~al.}(2018)\citenamefont
  {C{\'a}rdenas}, \citenamefont {Juli{\'e}},\ and\ \citenamefont
  {Deruelle}}]{Cardenas:2017chu}%
  \BibitemOpen
  \bibfield  {author} {\bibinfo {author} {\bibfnamefont {M.}~\bibnamefont
  {C{\'a}rdenas}}, \bibinfo {author} {\bibfnamefont {F.-L.}\ \bibnamefont
  {Juli{\'e}}},\ and\ \bibinfo {author} {\bibfnamefont {N.}~\bibnamefont
  {Deruelle}},\ }\href {https://doi.org/10.1103/PhysRevD.97.124021} {\bibfield
  {journal} {\bibinfo  {journal} {Phys. Rev. D}\ }\textbf {\bibinfo {volume}
  {97}},\ \bibinfo {pages} {124021} (\bibinfo {year} {2018})},\ \Eprint
  {https://arxiv.org/abs/1712.02672} {arXiv:1712.02672 [gr-qc]} \BibitemShut
  {NoStop}%
\bibitem [{\citenamefont {Guo}\ \emph {et~al.}(2026)\citenamefont {Guo},
  \citenamefont {Myung},\ and\ \citenamefont {Shao}}]{Guo:2026fga}%
  \BibitemOpen
  \bibfield  {author} {\bibinfo {author} {\bibfnamefont {H.}~\bibnamefont
  {Guo}}, \bibinfo {author} {\bibfnamefont {Y.~S.}\ \bibnamefont {Myung}},\
  and\ \bibinfo {author} {\bibfnamefont {L.}~\bibnamefont {Shao}},\ }\href@noop
  {} {\  (\bibinfo {year} {2026})},\ \Eprint {https://arxiv.org/abs/2609.20040}
  {arXiv:2609.20040 [gr-qc]} \BibitemShut {NoStop}%
\bibitem [{\citenamefont {Astefanesei}\ \emph {et~al.}(2019)\citenamefont
  {Astefanesei}, \citenamefont {Herdeiro}, \citenamefont {Pombo},\ and\
  \citenamefont {Radu}}]{Astefanesei:2019pfq}%
  \BibitemOpen
  \bibfield  {author} {\bibinfo {author} {\bibfnamefont {D.}~\bibnamefont
  {Astefanesei}}, \bibinfo {author} {\bibfnamefont {C.}~\bibnamefont
  {Herdeiro}}, \bibinfo {author} {\bibfnamefont {A.}~\bibnamefont {Pombo}},\
  and\ \bibinfo {author} {\bibfnamefont {E.}~\bibnamefont {Radu}},\ }\href
  {https://doi.org/10.1007/JHEP10(2019)078} {\bibfield  {journal} {\bibinfo
  {journal} {JHEP}\ }\textbf {\bibinfo {volume} {10}},\ \bibinfo {pages}
  {078}},\ \Eprint {https://arxiv.org/abs/1905.08304} {arXiv:1905.08304
  [hep-th]} \BibitemShut {NoStop}%
\bibitem [{\citenamefont {Juli{\'e}}\ and\ \citenamefont
  {Berti}(2019)}]{Julie:2019sab}%
  \BibitemOpen
  \bibfield  {author} {\bibinfo {author} {\bibfnamefont {F.-L.}\ \bibnamefont
  {Juli{\'e}}}\ and\ \bibinfo {author} {\bibfnamefont {E.}~\bibnamefont
  {Berti}},\ }\href {https://doi.org/10.1103/PhysRevD.100.104061} {\bibfield
  {journal} {\bibinfo  {journal} {Phys. Rev. D}\ }\textbf {\bibinfo {volume}
  {100}},\ \bibinfo {pages} {104061} (\bibinfo {year} {2019})},\ \Eprint
  {https://arxiv.org/abs/1909.05258} {arXiv:1909.05258 [gr-qc]} \BibitemShut
  {NoStop}%
\end{thebibliography}%
\end{document}